\documentclass[]{interact}

\usepackage{epstopdf}
\usepackage[caption=false]{subfig}
\usepackage{soul}
\usepackage{cite}
\usepackage[numbers,sort&compress,merge]{natbib}
\bibpunct[, ]{[}{]}{,}{n}{,}{,}
\renewcommand\bibfont{\fontsize{10}{12}\selectfont}

\theoremstyle{plain}

\theoremstyle{definition}

\theoremstyle{remark}

\usepackage{graphicx}
\usepackage{dcolumn}
\usepackage{bm}

\usepackage{xcolor} 
\usepackage{mathtools,algpseudocode,hyperref}
\usepackage{booktabs}  
\usepackage{tabularx}  
\usepackage{amsmath}   
\usepackage{lmodern}   
\usepackage{siunitx}   
\usepackage{caption} 
\usepackage{float}

\DeclarePairedDelimiter\ket{\lvert}{\rangle}
\DeclarePairedDelimiterX\braket[2]{\langle}{\rangle}{#1\,\delimsize\vert\,\mathopen{}#2}
\DeclarePairedDelimiter\expval{\langle}{\rangle}
\DeclarePairedDelimiter\abs{\lvert}{\rvert}

\newcommand{\be}{\begin{equation}}
\newcommand{\ee}{\end{equation}}

\newcommand{\ba}{\begin{array}}
\newcommand{\ea}{\end{array}}

\newcommand{\bea}{\begin{eqnarray}}
\newcommand{\ena}{\end{eqnarray}}

\newcommand{\beano}{\begin{eqnarray*}}
\newcommand{\enano}{\end{eqnarray*}}

\newcommand{\bei}{\begin{itemize}}
\newcommand{\eni}{\end{itemize}}

\newcommand{\bee}{\begin{enumerate}}  
\newcommand{\ene}{\end{enumerate}}

\def \ni  {\noindent}

\begin{document}


\title{Quantum propagation using Hagedorn wave packets: generalized scheme}

\author{
\name{Rabiou Issa \textsuperscript{a}\thanks{CONTACT R. Issa Email: issagoudouya@gmail.com}, 
Kokou Mawulonmi Robert Afansounoudji \textsuperscript{a}\thanks{
\qquad \qquad  \quad  K. M. R.Afansounoudji Email:   afankokourobert@gmail.com}, 
Komi Sodoga  \textsuperscript{a}\thanks{
\qquad \qquad  \quad K. Sodoga  Email: ksodoga@univ-lome.tg}, 
and 
 David Lauvergnat\textsuperscript{b}\thanks{
Corresponding author D. Lauvergnat Email: David.Lauvergnat@universite-paris-saclay.fr} 
 }
\affil{\textsuperscript{a}Laboratoire de Physique des Matériaux et des Composants à Semi-conducteurs (LPMCS) and Centre d’Excellence Régionale pour la Maîtrise de l’Electricité (CERME), University of Lomé, 01BP1515 Lomé, Togo; \\
\textsuperscript{b}Institut de Chimie Physique,  Bâtiment 349, 15 avenue Jean Perrin, CNRS UMR8000, Université Paris-Saclay,  91405 Orsay Cedex, France}
}

\maketitle

\begin{abstract}
In this study, we present a novel wave packet propagation method that generalizes the Hagedorn approach by introducing alternative primitive basis sets that are better suited to describe various physical processes. We can mix time-dependent (Hagedorn) and  time-independent basis sets, such as Fourier series, particle-in-a-box, and harmonic oscillator basis sets. Furthermore, our implementation can handle models with several electronic states, enabling the study of non-adiabatic processes. Instead of the time-dependent variational principle, our propagation scheme uses a three-step procedure (standard propagation, time-dependent parameter evaluation, and projection). This procedure relies on multidimensional integrations performed numerically with Gaussian quadrature, allowing us to avoid constraints on the form of the Hamiltonian operator.
We have implemented this numerical algorithm in Fortran code, and validated it by comparing it with standard propagation schemes on harmonic and anharmonic 2D-models. Finally, we have applied our method to compute the vibrational spectrum of the 6D-modified Hénon-Heiles model and we show that our scheme reproduces well the results obtained with the standard approach. As a perspective, we show that with our generalized Hagedorn wave packet method, we are able to study the non-adiabatic dynamics of the cis-trans retinal photoisomerization with a reduced 2D-model.
\end{abstract}


\section{Introduction} \label{Sec-intro}
\noindent Understanding the behavior of molecular systems requires quantum mechanics, which is essential for describing both their electronic structures and their nuclear motions. For example, the simulation of the dynamics of certain fundamental  processes, such as multiphoton excitations \cite{walker1994precision, rudenko2004correlated},  photodissociation \cite{engel2005calculated, pedersen2007crossed,sodoga2009photodissociation},  photo-isomerization \cite{hahn2000femtosecond,hahn2000quantum, marsili2020quantum, manathunga2017impact, Pereira2023} and charge transfer \cite{kimura1986charge,jain1987density,loreau2010ab}  requires to solve numerically the time-dependent Schrödinger equation (TDSE).  
However, studying these processes can be very challenging (see references \cite{Cigrang2025,Wu2025}. There are several possible approaches to study these processes, such as, semiclassical\cite{miller2001semiclassical,liu2007real,frantsuzov2003gaussian},  Mixed-Quantum Classical\cite{Tully1990,Coker1995,Kapral2016,Agostini2016,Runeson2023}, Gaussian Wave Packets\cite{martinez1996multi,worth2004novel,liu2011approach,curchod2018ab} and Wave Packet\cite{meyer1990multi,Wang2002,Manthe2008,Vendrell2011,Pereira2023} methods. \\

In the present study, we focus on quantum mechanical approaches to solve the TDSE using wave packet propagation techniques, so that restricting the studies to a temperature of 0 K. Additionally, the studies will be limited to systems of a finite-size, without infinite environments. Therefore, to solve the TDSE, we need to determine the time evolution of a wave packet, $\Psi$. Usually, the propagation is divided into small constant time intervals, $\Delta t$. Hence, the wave packet, $\ket{\Psi\left( t + \Delta t\right)}$, at $t+\Delta t$ is obtained from the wave packet at $t$, $\ket{\Psi\left( t \right)}$, as follows:
\begin{eqnarray}
	\ket{\Psi\left( t + \Delta t\right)} &=& \exp\big[ -\imath/\hbar \cdot \hat{H}\Delta t\big]\ket{\Psi\left( t\right)}.
\end{eqnarray}
where $\exp\big[ -\imath/\hbar \cdot \hat{H}\Delta t\big] $ is the evolution operator when the Hamiltonian is time-independent. Without being exhaustive, to solve the TDSE numerically, one can  use standard methods \cite{leforestier1991comparison}: 
(i) the split operator method \cite{feit1982solution}, 
(ii) decomposition of the evolution operator into polynomials (usually Chebychev) \cite{tal1984accurate}, 
(iii) Taylor expansion of the evolution operator \cite{lauvergnat2007simple}.
However, some of the above-mentioned methods\cite{leforestier1991comparison,feit1982solution,tal1984accurate,lauvergnat2007simple} have drawbacks related to the number of nuclear degrees of freedom in the system. Indeed, when a direct-product is used, the size of the basis (or grid) scales exponentially with the number of degrees of freedom. In addition, the basis must accurately describe the evolution of the wave packet over time. Thus, if the wave packet delocalizes, the basis describing it should be very large. Similarly, if the center of the wave packet evolves strongly while remaining localized, the basis must also be very large. 
Furthermore, non-adiabatic processes present significant challenges to simulations. These transitions between electronic states often involve a topological feature known as a conical intersection.
At this intersection, nuclear and electronic motions become intertwined by so-called non-adiabatic coupling, where quantum effects play a crucial role in the dynamics of the nuclear motion. 

To solve the problem when the center of the wave packet evolves strongly, it is possible to represent the wave packet in a basis set (and grid) that evolves over time. More precisely, some basis set parameters change variationally during the propagation. Several wave packet propagation techniques have been developed, based on this idea of a time-evolving basis set (or grid). Without being exhaustive, one can mention: 
(i) The semiclassical approach introduced by Heller\cite{heller1976time, huber1987generalized,drolshagen1983time,
	vanivcek2021ab}
in which a single Gaussian evolves over time. 
(ii)  The approaches in which the wave packet is described by a sum of Gaussians evolving over time \cite{worth2004novel,martinez1996multi,curchod2018ab}.
(iii) In the references \cite{sielk2009quantum,Choi2019}, the wave packet is represented 
on a grid that changes over time according to the evolution of the wave packet.
(iv) A widely adopted solution in this context is the multi-configuration time-dependent Hartree (MCTDH) method \cite{meyer1990multi,hammerich1990quantum, meyer2003quantum,Pelaez2017}. This approach represents the wavefunction as a small direct product of time-dependent basis sets. This method allows the propagation of quantum wave packets with a few dozen of degrees of freedom, or even larger with the multilayer version of MCTDH \cite{Wang2002,Manthe2008,Vendrell2011,Shi2023}. However, this approach has its limitations. Both the kinetic and the potential components of the Hamiltonian must be expressed as sums of products of one-dimensional terms, which is not always an easy task.
(iv) George Hagedorn introduced the approach now known as Hagedorn wave packets\cite{hagedorn1981semiclassical, hagedorn2006ac, hagedorn1998raising,faou2009computing,lasser2020computing,zhang2024single}, i.e.  products of complex Gaussians with polynomials,  also known as a prefactor polynomial \cite{dietert2017invariant, vanivcek2024hagedorn, ohsawa2019hagedorn}, that form an orthonormal basis in $L^2$ space. Certain parameters (center, width and momentum) vary variationally during the propagation.
Detailed presentations of these functions can be found in the literature \cite{hagedorn2013minimal, ohsawa2013symplectic, gradinaru2014convergence, lasser2014hagedorn, li2014improved, vanivcek2024hagedorn, ohsawa2015symmetry, punovsevac2016dynamics}.
However, the standard Hagedorn scheme is not general. Indeed, it does not allow the use of other types of primitive basis sets, such as the particle-in-a-box or the Fourier basis sets, which are often used to model torsional motions in molecules. Furthermore, the standard Hagedorn scheme is restricted to a single electronic state. \\
This work aims to  present a generalized Hagedorn wave packet propagation scheme that balances flexibility and numerical efficiency. Unlike the standard Hagedorn method, which relies on a specific basis set tied to a single electronic state, our approach:
	(i) combines time-dependent (Hagedorn) and time-independent basis sets (e.g., Fourier and particle-in-a-box) to better model diverse processes, such as torsional motions.
	(ii)  handles multi-state non-adiabatic dynamics via a diabatic representation, critical for studying processes involving conical intersections.
	(iii)  avoids variational constraints through a three-step algorithm: standard propagation, parameter update, and projection. This enables broader Hamiltonian compatibility. \\
More precisely, we provide a modern Fortran code (freely available on Github \cite{ISSA_2025}) implementing this three-step propagation technique that generalizes the multidimensional Hagedorn basis set and works with other types of primitive basis sets.

The remainder of the paper is structured as follows. Section 2 provides a detailed presentation of the relevant theoretical methods. In Section 3, we present the results of tests conducted on 2D models with a single electronic surface. Section 4 focuses on applications, including the 6D Hénon-Heiles model. Finally, Section 5 summarizes the key points and discusses the results of this work. Furthermore, as perspective to non-adiabatic dynamics, the isomerization of retinal is considered as a 2D model with two electronic surfaces.
\section{Methods} \label{Sec-Meth}

\noindent As mentioned in the introduction, the aim of the present study is to propose a propagation strategy with a time-dependent basis set, closely related to the Hagedorn wave packet propagation scheme. However, we will also compare this new approach against a standard approach with a time-independent basis set. \\
\noindent  
For a molecular system composed of $nc$ coordinates, $\mathbf{Q} \left( \mathbf{Q}=\left\lbrace  q_1, q_2 \cdots q_{nc}\right\rbrace \right) $, the total wave packet, $\ket{\Psi\left( t\right) }$, is expanded, as usual, on $ne$ electronic diabatic states, $\ket{e}$, as follows: 
	\begin{eqnarray} \label{Eq-PsiTot}
		\ket{\Psi\left( t\right) } &=& \sum_{e=1}^{ne}\ket{\psi^{(e)}\left( t\right) }  \ket{e}
	\end{eqnarray}
	where, $\ket{\psi^{(e)}\left( t\right) }$, represents the nuclear wave packet, $\psi^{(e)}\left( \mathbf{Q},t\right) $, on the $e^{th}$ electronic diabatic state. This electronic component can be expanded on a  multidimensional basis and written as:
	\begin{equation} \label{Eq-Psi_e}
		\Psi^{(e)}\left( {\bf Q},t\right)  = \sum_I^{NB} \tilde{C}_I^t \cdot B_I\left( \mathbf{Q},\mathbf{\Lambda}\right) , 
	\end{equation} 
	where $ B_I\left( \mathbf{Q, \Lambda}\right) $ is the $I^{th}$ \emph{time-independent} basis function and $\mathbf{\Lambda} \equiv \left\lbrace \lambda_1, \ldots, \lambda_k, \\ \ldots, \lambda_{nc} \right\rbrace$  are \emph{time-independent}  basis parameters (such as, the center and the scaling factor of the Harmonic Oscillator, HO, basis set). $\tilde{C}_I^t$ is the $I^{th}$ \emph{time-dependent} coefficient, and $NB$ the size of the basis set.

In the present study, the standard propagation is performed with a simple but efficient Taylor expansion\cite{Mercouris1994,lauvergnat2007simple} of the wave packet around $t$. The wave packet at $t+\Delta t$ is obtained as follows:
\begin{eqnarray} \label{Eq-PsiTaylor}
\ket{\Psi(t+\Delta t)} &=& \sum_{\ell=0}^{\ell_{max}} \frac{\Delta t^\ell}{\ell!} \ket{\Psi^{(\ell)}\left( t\right) },
\end{eqnarray}
where in the equation \ref{Eq-PsiTaylor}, $\ket{\Psi^{(\ell)}\left( t\right) }$ is the $\ell^{th}$ derivatives of the wave packet at $t$. It can be obtained recursively from the time-dependent Schrödinger equation:
\begin{eqnarray} \label{Eq-PsiDeriv}
\ket{\Psi^{(\ell+1)}(t)} &=& -\imath/\hbar \cdot \hat{H}\ket{\Psi^{(\ell)}\left( t\right) }.
\end{eqnarray}
Remarks: $\ket{\Psi^{(0)}\left( t\right) }$ ($\ell=0$) is simply the wave packet at time $t$. The previous relation (Eq. \ref{Eq-PsiDeriv}) is correct only when the Hamiltonian is time-independent. When it is not the case, time derivatives of the Hamiltonian must be taken into account\cite{lauvergnat2007simple,Mercouris1994}.
Furthermore, the order, $\ell_{max}$, of the Taylor series is obtained when the norm of the $\ell_{max}{}^{th}$ term, $\frac{\Delta t^{\ell_{max}}}{\ell_{max}!} \ket{\Psi^{(\ell_{max})}(t)}$, is smaller than a threshold (here $10^{-20}$). This guarantees that the wave packet is propagated with high accuracy, so that its norm and energy are numerically conserved over time.

\subsection{Wave packet with time-dependent basis set} 
\noindent When the basis set is time-dependent, the wave packet expansion is similar to Eq. \ref{Eq-Psi_e}, except that the basis parameters, $\mathbf{\Lambda}_{t}$, are time-dependent. Then this expansion becomes:
\be \label{Eq-PsiNuc}
\psi^{(e)}\left( \mathbf{Q},t\right)   = \sum_{I=1}^{NB} C_I^t \cdot B_{I}\left( \mathbf{Q},\mathbf{\Lambda}_{t} \right).
\ee
In the previous equation, (Eq. \eqref{Eq-PsiNuc}), $B_{I}\left( \mathbf{Q},\mathbf{\Lambda}_{t} \right)$ is the $I^{th}$ \emph{time-dependent} multidimensional basis function, the $C_{I}^t$ is its corresponding coefficient. The $\mathbf{\Lambda}_{t} $ are the \emph{time-dependent} parameters of the basis set common to all electronic states, where $\mathbf{\Lambda}_{t} =\left\lbrace \lambda^{(t)}_{1}, \cdots, \lambda^{(t)}_{k}, \cdots \lambda^{(t)}_{nc}\right\rbrace $. Each basis function is expressed as a product of 1D-primitive basis functions, $b^{k}_{i_{k}}\left( q_{k},\lambda^{(t)}_{k} \right)$, as follows:
\begin{eqnarray} \label{Eq-nDBasis}
B_{I}\left( \mathbf{Q},\mathbf{\Lambda}_{t}\right)  &=& \prod_{k=1}^{nc} b^{k}_{i_{k}}\left( q_{k},\lambda^{(t)}_{k}\right).
\end{eqnarray}
In the previous equation (Eq. \ref{Eq-nDBasis}), $I$ is corresponding to a multidimensional index, $\left\lbrace   i_{1},\cdots,i_{k},\cdots,i_{nc}\right\rbrace $, associated with the 1D-primitive basis functions. The total number of functions, $NB$, is expressed as $\prod\limits_{k=1}^{nc} nb_{k}$ for a direct-product basis set, where $nb_k$ is the number of functions of the $k^{th}$ primitive basis set.
The $\lambda^{(t)}_{k}$ are the time-dependent parameters of the $k^{th}$ primitive basis set and are particularized according to its type: 
    (i) for the \emph{Fourier basis set} there are no time-dependent parameters 
    (ii) for the \emph{sine}  or the \emph{particle-in-box basis set}, the parameters (the domain range), $\lambda_{k} =\left\lbrace  A_{k},B_{k} \right\rbrace$, are time-independent. Although, they can also be implemented as time-dependent parameters.
    (iii) for the Hagedorn primitive basis set, the time-dependent parameters are particularized in $\lambda^{(t)}_{k} = \left\lbrace  \alpha^{(t)}_{k},q^{(t)}_{k},p^{(t)}_{k} \right\rbrace $ where $\alpha^{(t)}_{k}$ is related to the basis set width and $q^{(t)}_{k}$, $p^{(t)}_{k}$ represent the center, and the momentum of the Hagedorn basis set, respectively.
More precisely, the expression of the Hagedorn primitive basis function, $b^{k}_{i_{k}}\left( q_{k},\lambda^{(t)}_{k}\right)$ (or more concisely  $b^{k}_{i_{k}}\left( q_{k},t \right)$), is given by:
\begin{eqnarray} \label{Eq-HagBasis}
b^{k}_{i_{k}}\left( q_{k},t \right) 
&\equiv & 
 b^{k}_{i_{k}}\left( q_{k}, \lambda^{(t)}_{k} \right)  \cr
& = &  N\left( \lambda_{k}^{(t)}\right) 
 \cdot H_{i_{k}-1}\left(  x_{k} \right) 
  \times  \exp\left( -\frac{\alpha^{(t)}_{k}}{2}\Delta q_{k} ^{2}+\imath \cdot p^{(t)}_{k} \Delta q_{k} \right)
\end{eqnarray}
where $\alpha^{(t)}_{k}$ is a complex number ($\alpha^{(t)}_{k}=a^{(t)}_{k} + \imath \cdot b^{(t)}_{k}$) where $a^{(t)}_{k}$ and $b^{(t)}_{k}$ are its real and imaginary parts, respectively.
In Eq \eqref{Eq-HagBasis}, $ \Delta q_{k} = q_{k}-q^{(t)}_{k} $  and  $ x_{k} = \sqrt{a^{(t)}_{k}}\Delta q_{k}$. The square root of $a^{(t)}_{k}$ is a scaling factor and it is related to the width of the Gaussian part of the basis set. Therefore, $a^{(t)}_{k}$ is positive. Furthermore, $N\left( \lambda_{k}^{(t)}\right)$ is a normalization parameter that depends only on $a^{(t)}_{k}$ and $H_{i_{k}}\left(  x_{k} \right)$ is the usual normalized Hermite polynomial of degree $i_{k}$.
\noindent It is worth noting that: (i) The Hagedorn and Harmonic Oscillator (HO) basis sets are similar, except that the imaginary contributions of the Hagedorn basis set are absent in the HO ones. (ii)
Some parameters can be excluded for the time-dependent basis. For instance, the imaginary part $b^{(t)}_{k}$ of $\alpha^{(t)}_{k}$ or $p^{(t)}_{k}$. This does not prevent an exact propagation, since the missing contributions are transferred to the $C_{I}^t$ coefficients if the basis set is large enough. (iii) The Heller approach \cite{heller1976time,heller1981semiclassical,heller1975time} can be cast in the Hagedorn scheme where only one basis set is present ($nb_{k}=1$ and hence, $NB=1$). (iv) In the standard Hagedorn or Heller schemes, the time-dependent parameter $\bm{\alpha}^{(t)}$ is a full complex $nc \times nc$ matrix. In our scheme, this matrix is diagonal and the diagonal elements are the $\alpha^{(t)}_{k}$. This approximation does not prevent an exact propagation (the missing coupling contributions are transferred to the $C_{I}^t$ coefficients) and was made to facilitate the standard grid-to-base transformation required in our scheme. (v) With our scheme, we can mix time-dependent and time-independent basis sets, and it also works with several electronic states.

\subsection{Propagation strategy}

\begin{figure}[hbtp]
\begin{center}
\includegraphics[width=14cm]{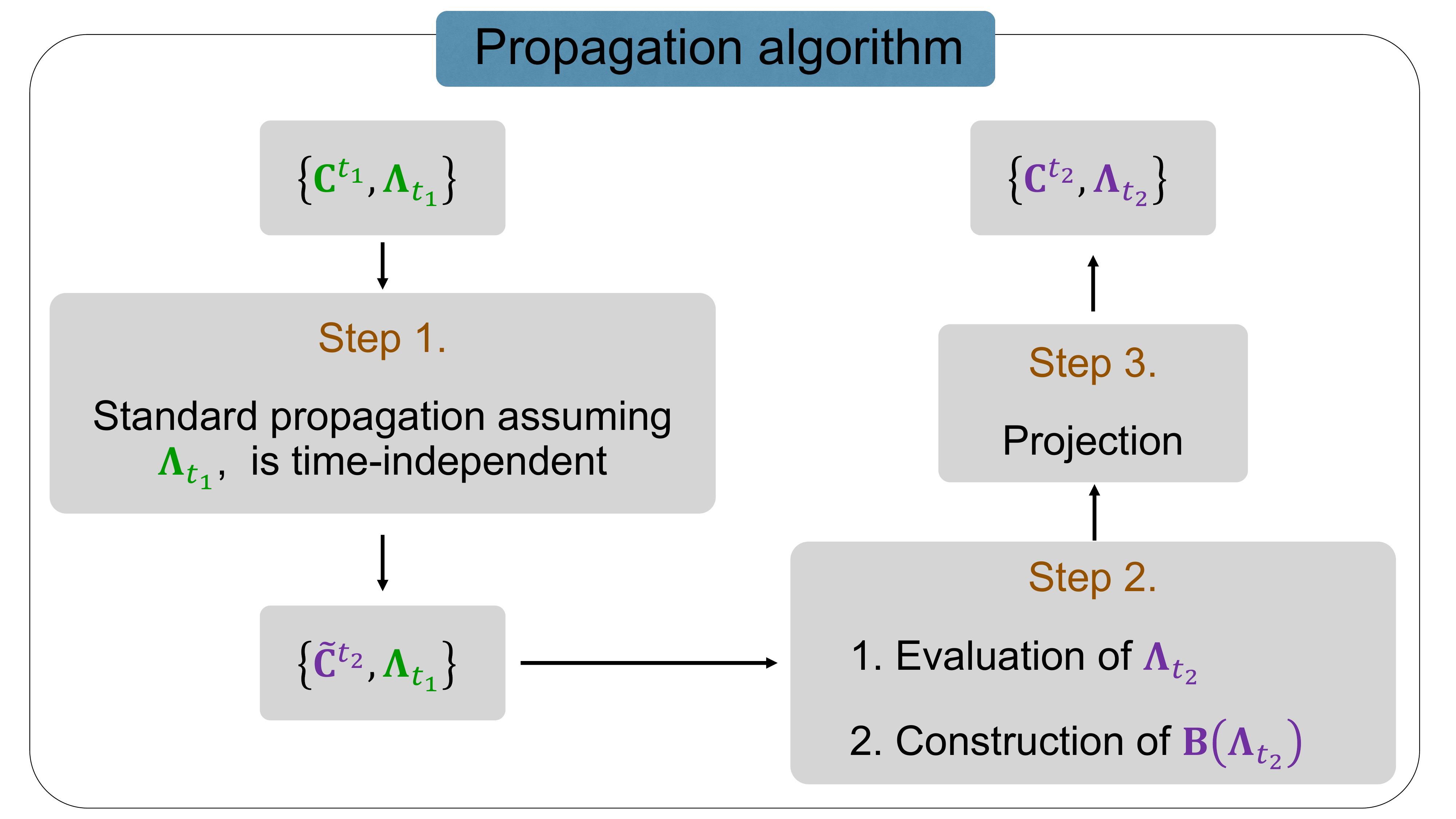}
\caption{\textit{Propagation algorithm from $t_1$ to $t_2$ in three steps. \emph{Step 1}: Standard propagation with a time-independent basis set (parameters, $\mathbf{\Lambda}_{t_1}$ at $t_1$). \emph{Step 2}: Evalutation of the basis parameters, $\mathbf{\Lambda}_{t_2}$, and basis set construction at $t_2$. \emph{Step 3}: Projection of the wave packet at $t_2$ expanded on the basis set in $t_1$ into the the basis set in $t_2$. re-normalization if required.}}
\label{Fig_algo}
\end{center}
\end{figure}

\noindent The key point of our Hagedorn approach lies in the way we propagate the  time-dependent parameters $ \left\lbrace  \mathbf{C}^{t}, \mathbf{\Lambda}_{t}\right\rbrace  $, where $\mathbf{\Lambda}_{t}$ are the basis set parameters and $\mathbf{C}^{t}=\left\lbrace C_{1}^t, \cdots C_{NB}^t\right\rbrace $ are the wave packet coefficients.
Indeed, the usual way to propagate the Hagedorn wave packet or other approaches (Heller \cite{huber1987generalized,drolshagen1983time,moghaddasi2023high}, Variational Multi-Configuration Gaussian\cite{lasorne2011excited,Lasorne2014,lasorne2006direct,
lasorne2007direct,
mendive2012towards,allan2010straightforward,araujo2009molecular,lasorne2008controlling,araujo2010controlling,richings2015quantum}, MCTDH \cite{beck2000multiconfiguration,MCTDHWebsite}) from time $t_1$ to time $t_2$ is to use the time-dependent variational principle. However, we use a different strategy with three steps (fig. \ref{Fig_algo}):
\begin{enumerate}
\item \emph{Standard propagation}: In this step (See fig. \ref{Fig_algo}), we perform a standard propagation technique (here, the Taylor propagation scheme, but other schemes could be used\cite{leforestier1991comparison}) assuming a time-independent basis, $\mathbf{B}\left( \mathbf{\Lambda}_{t_1}\right)$, to obtain an intermediate wave packet at $t_2$ :

\begin{equation}
\label{Eq-psiInter}
\Psi\left(\bm{Q},t_2\right)  = \sum_{I=1}^{NB} \tilde{C}^{t_2}_{I} \cdot B_{I}\left(\bm{Q}, \bm{\Lambda}_{t_1}\right).
\end{equation}
%
We obtain a new set of $ \left\lbrace  \mathbf{\tilde{C}}^{t_2}, \mathbf{\Lambda}_{t_1} \right\rbrace $. It is important to note that $\mathbf{\Lambda}_{t_1}$ is still at time $t_1$ and the wave packet coefficients, $\mathbf{\tilde{C}}^{t_2}$, are not the final ones. Furthermore, and depending on the propagation technique used, the time step ($ \Delta t=t_2 - t_1 $) can be relatively large.
\item \emph{New basis set}: Then we build the new basis set, $\mathbf{B}\left( \mathbf{\Lambda}_{t_2}\right)$, by computing the parameters $\mathbf{\Lambda}_{t_2}$, at time $t_2$ (See fig. \ref{Fig_algo}). More precisely, the parameters, $q^{(t)}_{k}$, $p^{(t)}_{k}$ and $\alpha^{(t)}_{k}$ (for $k=1, \cdots nc$), are computed from the expectation values of operators assuming the wave packet has a Gaussian shape (Heller-like wave packet):
{
\begin{eqnarray}
q^{(t)}_{k}  &=&  \expval*{\Psi\left( t\right) \middle| q_{k} \middle| \Psi\left( t\right) }, \label{Eq-expect_qk} \\
p^{(t)}_{k}  &=& -\imath   \expval*{\Psi\left( t\right) \middle| \partial q_{k} \middle| \Psi\left( t\right) } \label{Eq-expect_pk} \\
a^{(t)}_{k}  &=& \left[ 2\left( \expval*{\Psi\left( t\right) \middle| q_{k}{^2} \middle| \Psi\left( t\right) }-\left( q^{(t)}_{k}\right)^{2} \right) \right]^{-1},  \label{Eq_ak} \\ 
b^{(t)}_{k} &=& a^{(t)}_{k}\left[ 2p^{(t)}_{k}q^{(t)}_{k} 
+ \imath\left( 1 + 2 \expval*{\Psi\left( t\right) \middle| q_{k}\frac{\partial}{\partial {q_{k}}} \middle| \Psi\left( t\right) } \right) \right].  \label{Eq_bk}
\end{eqnarray}
}
 In the previous equations, $q^{(t)}_{k}$ and $p^{(t)}_{k}$ are, respectively, the expectation values of the position operator, $q_k$, (Eq. \ref{Eq-expect_qk}) and momentum operator, $\hat{p}_k$, (Eq. \ref{Eq-expect_pk}). As mentioned above, the parameter $a^{(t)}_{k}$ (Eq. \ref{Eq_ak}) is related to the width or the variance of the wave packet and can be computed from the expectation values of the ${q_k}^2$ and $q_k$ operators. There are several ways to compute $b^{(t)}_{k}$, but some of them are not numerically stable. Therefore, in the present implementation (Eq. \ref{Eq_bk}), we used the expectation value of the $q_{k}\frac{\partial}{\partial {q_{k}}}$ operator and the computed expressions of $q^{(t)}_{k}$ and $p^{(t)}_{k}$.
It is worth mentioning, these time-dependent parameters are not optimal (except for the Heller wave packet), but this does not prevent accurate propagation.
\item \emph{Projection}: 
	After, the standard propagation (the first step), the wave packet is known at time $t_2$ and as a projection onto the basis set at time $t_1$ (the old basis set, $\mathbf{B}\left( \mathbf{\Lambda}_{t_1}\right)$). The wave packet expansion is characterized by the coefficients, $\mathbf{\tilde{C}}^{t_2}$, and the time-dependent basis parameters, $\mathbf{\Lambda}_{t_1}$ (Eq. \ref{Eq-psiInter}). However, this wave packet must be projected onto the newly constructed basis set $\mathbf{B}\left( \mathbf{\Lambda}_{t_2}\right)$ at time $t_2$ obtained in the second step:
	\begin{eqnarray}
	\label{Eq-psifinal}
	\Psi(\bm{Q},t_2) & = &\sum_{I=1}^{NB} C^{t_2}_{I} \cdot B_{I}\left( \bm{Q}. \bm{\Lambda}_{t_2}\right)
\end{eqnarray}
Then, this wave packet expansion is characterized by the coefficients, $ \mathbf{C}^{t_2}$, and the time-dependent basis parameters, $\mathbf{\Lambda}_{t_2}$. 
It is then assumed that the two expansions, Eq. \ref{Eq-psiInter} and Eq. \ref{Eq-psifinal}, are equal:
		\begin{eqnarray}
		\label{Eq-psiInterEQfinal}
		\sum_{J=1}^{NB} C^{t_2}_{J} \cdot B_{J}\left( \bm{Q}, \bm{\Lambda}_{t_2}\right) & = &  \sum_{I=1}^{NB} \tilde{C}^{t_2}_{I} \cdot B_{I}\left(\bm{Q}, \bm{\Lambda}_{t_1}\right).
	\end{eqnarray}
	Then, to obtain the $J^{th}$ new coefficient, $C^{t_2}_{J}$, the previous equation (Eq \ref{Eq-psiInterEQfinal}) is projected onto the new basis function, $B_{J}\left( \bm{Q}, \bm{\Lambda}_{t_2}\right)$:
	\begin{eqnarray} \label{Eq-projectcoef}
		C^{t_2}_{J} &=& \sum_{I=1}^{NB} \braket*{B_{J}\left( \bm{Q}, \bm{\Lambda}_{t_2}\right) }{B_{I}\left( \bm{Q}, \bm{\Lambda}_{t_1}\right) } \tilde{C}^{t_2}_{I}.
	\end{eqnarray}
It is important to note that, the equality assumption (Eq. \ref{Eq-psiInterEQfinal}) only holds true when the basis set size is infinite in the second expansion (Eq. \ref{Eq-psifinal}). However when the basis set size is equal to $NB$, the numerical values of the new coefficients, $\mathbf{C}^{t_2}$, obtained from the projection  (Eq. \ref{Eq-projectcoef}), are still numerically exact. Furthermore, if the number of terms in the sum is be too small  (Eq. \ref{Eq-psifinal}), the wave packet norm may not be conserved (see solutions below to overcome this difficulty).
\noindent This equation (Eq. \ref{Eq-projectcoef}) shows that the projection is performed on a multidimensional basis. However, numerically, it can be achieved sequentially using partial projections (for one 1D-basis set, see the Appendix \ref{WPProject} for more details). It is important to note that this sequential transformation is possible because \(\alpha^{(t_2)}_k\) is considered to be a diagonal matrix. When, it is not the case, one can use an algebraic method as proposed by Van{\'{i}}{\v{c}}ek et al. \cite{vanivcek2024hagedorn}.
\end{enumerate}
\quad \newline
\quad \newline
Some comments are in order: in the first step, the wave packet norm and energy can be conserved with some high numerical accuracy, and with some propagation techniques (such as Lanczos) they are rigorously conserved.\cite{leforestier1991comparison} However, during the third step, the  wave packet norm may not be conserved if the basis set size, $NB$, is not large enough or if the time step, $\Delta t$, is too large. Several strategies can be used to overcome this difficulty: (i) increase the basis set size, (ii) reduce the time step, (iii) renormalize the wave packet with the norm after the first step.
Furthermore, in our implementation, we want to test several options regarding the number of time-dependent basis parameters. More precisely, $b^{(t)}_{k}$ (the imaginary part of $\alpha^{(t)}_{k}$) and the term with $p^{(t)}_{k}$ can be present or absent in the time-dependent basis parameters.
For the figures and the tables, the following notations are used:
\begin{itemize}
\item For the Hagedorn propagation scheme, the keyword HAG is used, then the number of 1D-basis function is added (assuming this number is the same for all dimensions), then two times a letter to specify when $b^{(t)}_{k}$ and $p^{(t)}_{k}$ are present, T (for true) or not F (for false), respectively. Finally another letter is added to indicate whether the re-normalization at the end of step 3 is used (T) or not (F). For instance, HAG5TTF means: Hagedorn propagation scheme with 5 basis functions for each dimension and keeping the time dependent parameters, $b^{(t)}_{k}$ and $p^{(t)}_{k}$. In addition, the re-normalization is not used.
\item For propagation with a time-independent basis set, the keyword, STD is used and then the number of 1D-basis functions is added. STD10 means: Standard propagation scheme with 10 basis functions for each dimension.
\end{itemize}
\section{Computational tests}
\noindent In this section, the accuracy of our approach is tested on two 2D-models (Eq. \ref{Eq-mod}) by comparing a wave packet obtained by the Hagedorn propagation scheme with another one converged with high accuracy obtained by a standard propagation method. More specifically, the three steps of our approach are tested. For the 2D-model ($nc=2$), the Hamiltonian model is given by:
\be \label{Eq-mod}
\hat{H}\left(  \mathbf{Q} \right) =  - \frac{\hbar^2}{2m} \sum_{k=1}^{nc}
\frac{\partial^2}{\partial {q_{k}}^2} + V\left(  \mathbf{Q} \right)
\ee 
\noindent where $m$ is the mass (here $m=1$) and $V\left(  \mathbf{Q} \right)$ is the potential. 
The initial wave packet is assumed to be a Gaussian and is defined as follows:
\begin{eqnarray} \label{Eq-Psi0}
\psi^{(e_0)}_0( \mathbf{Q}) = \prod_{k=1}^{nc} \exp \left( -\frac{a^{0}_{k}}{2}\Delta_{k} ^{2} +
\imath \cdot p^{0}_k \Delta_{k} \right), \quad
\Delta_{k} = q_{k} - q^{0}_k,
\end{eqnarray}
where, $e_0$ is the initial electronic state (in this model, $e_0=1$), $p^{0}_k$ and $a^{0}_{k}$ are, respectively, the initial momentum and a parameter related to the width of the Gaussian along the $k^{th}$ coordinates ($q_{k}$).
\subsection{2D-Harmonic Hamiltonian}
\noindent In this section, we will use a simple 2D-quadratic potential defined as follows: 
\begin{eqnarray} \label{Eq-2DHarm}
V\left( \mathbf{Q} \right) &=& \frac{1}{2} \left( {q_{1}}^2+{q_{2}}^2 \right).
\end{eqnarray} 
With this 2D-harmonic model and a Gaussian initial wave packet (Eq.\ref{Eq-Psi0}), the wave packet must remain Gaussian during propagation, which is equivalent to the Heller propagation scheme \cite{heller1976time,heller1975time}. This model enables us to check the influence of the time step and the basis set size, $nb_k$ on the propagation accuracy.
For this model, the Hagedorn basis set parameters and the initial wave packet parameters are given in the table (\ref{tab-HAGbasis-WP0-2D}) and have been chosen so that the parameters associated with the first coordinate ($k=1$) evolve during the propagation, while those associated with the second coordinate ($k=2$) remain constant during the propagation. Furthermore, the number of grid points is determined by adding $5$ to the number of basis functions, such that $nq_k = nb_k + 5$. This choice ensures high accuracy during numerical integration using Gaussian quadrature.
 
 \begin{table}[H]
 	\captionsetup{width=\textwidth}
 	\caption{\textit{Hagedorn and HO basis set parameter values: $q_k^{(0)}$, $p_k^{(0)}$, and $\alpha_k^{(0)}$ (columns 2--4). Initial Gaussian wave packet parameters (Eq.~\ref{Eq-Psi0}, with $e_0 = 1$): $q_k^0$, $p_k^0$, and $a_k^0$ (columns 5--7)}}
 	\label{tab-HAGbasis-WP0-2D}
 	\centering
 	\begin{tabularx}{\textwidth}{@{}c *{3}{>{\centering\arraybackslash}X} *{3}{>{\centering\arraybackslash}X}@{}}
 		\toprule
 		& \multicolumn{3}{c}{Hagedorn and HO basis sets} & \multicolumn{3}{c}{Initial wave packet} \\ 
 		\cmidrule(lr){2-4} \cmidrule(lr){5-7}
 		$k$ & $q_k^{(0)}$ & $p_k^{(0)}$ & $\alpha_k^{(0)}$ & $q_k^0$ & $p_k^0$ & $a_k^0$ \\ 
 		\midrule
 		1 & 2.0 & 0.0 & $(1.2, 0.0)$ & 2.0 & 0.0 & 1.2 \\
 		2 & 0.0 & 0.0 & $(1.0, 0.0)$ & 0.0 & 0.0 & 1.0 \\ 
 		\bottomrule 
 	\end{tabularx}
 \end{table}

\begin{figure}[hbtp]
\begin{center}
\includegraphics[width=14cm]{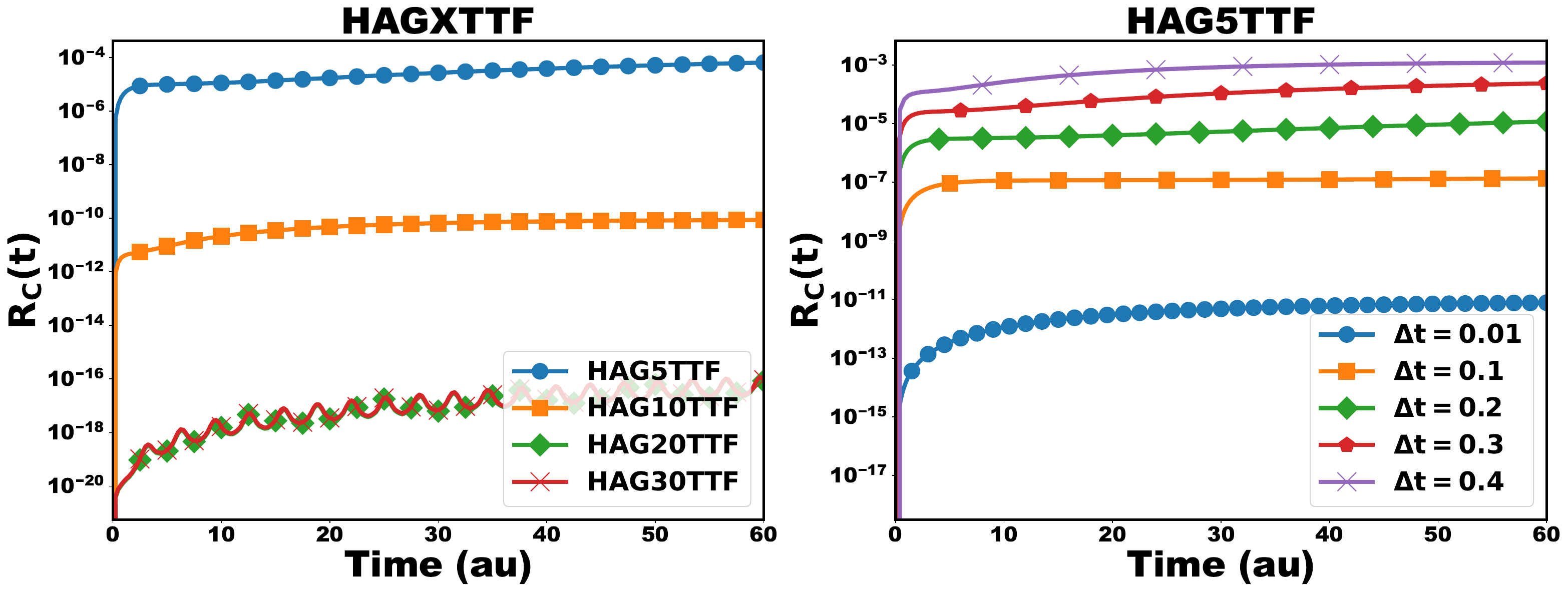}
\caption{\textit{ Time evolution of the wave function residual coefficient $R_C\left( t\right) $. The left panel presents $R_C\left( t\right) $ as a function of the basis size $nb_k$ for a fixed time step $\Delta t = 0.25\,au $, while the right panel shows $R_C\left( t\right) $ as a function of the time step $\Delta t$ with a fixed basis size $nb_k = 5$. The residual coefficient $R_C\left( t\right) $ quantifies the wave function deviation from the Gaussian shape at time $t$.}} \label{fig_RC-2DHarm}
\end{center}
\end{figure}

\ni First, we perform the overall analysis of the wave packet along time to check the numerical stability of our algorithm. As mentioned above, and since the wave packet remains Gaussian along time with this harmonic model, the wave packet must have a basis expansion where only the first coefficient, $C_1^t$, is non-zero and the other coefficients are zero ($C_I^t=0, (I>1)$). Since the wave packet is normalized, $\abs{C_1^t}=1$.
Therefore, the residual coefficient error $ R_C\left( t\right)  = \sum\limits_{I=2}^{NB} \left| C_{I}^{t}\right| ^2 $ and the norm conservation error, $\Delta N_3\left( t\right)  = \abs{N_{3}(0)-N_{3}\left( t\right) }$, allow to measure the wave packet deviation from its Gaussian shape. $N_3(t)$ denotes the norm of the wave packet at time $t$ after the three-step propagation scheme. 
%
\begin{figure}[hbtp]
\begin{center}
\includegraphics[width=14cm]{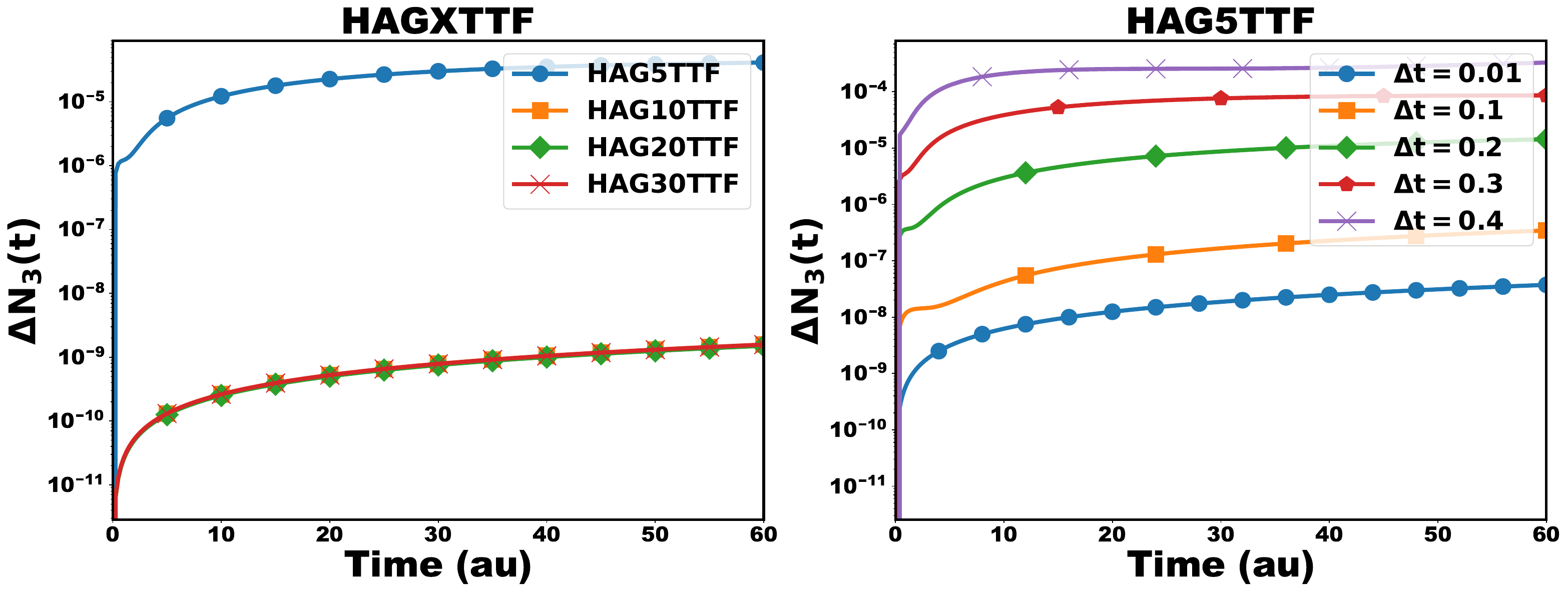}
\caption{\textit{ Time evolution of the wave packet norm deviation $\Delta N_{3}\left( t\right) $. 
The left panel shows $\Delta N_{3}\left( t\right) $ as a function of the basis size $nb_k$, for a fixed time step $\Delta t = 0.25\,au$. The right panel displays $\Delta N_{3}\left( t\right) $ as a function of the time step $\Delta t$, with a fixed basis size $nb_k = 5$. The quantity $\Delta N_{3}(t) \equiv \left| N_{3}\left( t\right)  - N_{3}\left( t_0\right) \right| $ measures the deviation between the instantaneous norm $N_{3}(t)$ and the initial wave packet norm $N_{3}\left( t_0\right) $}.} \label{fig_N3-2DHarm}
\end{center}
\end{figure}
%
\ni The figure \eqref{fig_N3-2DHarm} shows the time evolution of the norm conservation error, $\Delta N_3(t)$, as a function of the number of basis functions, $nb_k$, in each dimension with a fixed time step of $0.25\,au$ (left panel) and as a function of the time step, $\Delta t$ with $nb_k=5$. 
As expected, the norm conservation error becomes smaller  as $nb_k$ increases. For $nb_k=10$ or larger, the error is around $10^{-9}$.
Furthermore, with only five basis functions in each dimension, the norm error tends to zero as the the time step decreases. For $\Delta t=0.01\,au$, the error is approximately $4.10^{-8}$.
The residual coefficient error, $R_C(t)$, shows similar trends as $nb_k$ increases or as $\Delta t$ decreases (see Fig. \ref{fig_RC-2DHarm}). The fact that this error tends to zero shows that the wave packet remains Gaussian with time, since only the first coefficient, $C_1^t$, is non-zero. Furthermore, its absolute value must remain one, since the wave packet norm is one (Fig. \ref{fig_N3-2DHarm}).
The energy conservation error (not shown),  
$ \Delta E_{3}(t) = \abs*{	\expval*{\Psi\left( 0 \right)  \middle| \hat{H} \middle| \Psi\left( 0\right) }  -	\expval*{\Psi\left( 0 \right)  \middle| \hat{H} \middle| \Psi\left( t\right) } }$
, also presents a similar trend and with $\Delta t=0.01\,au$ and $nb_k=5$, this error is about $1.4\times10^{-12}\,au$.
%
\quad \newline
%
\ni Let us now analyze the three steps of our numerical scheme:
\begin{enumerate}
\item[(i)] The first step corresponds to a standard propagation using a wave packet Taylor expansion, for which the Taylor order is adjusted until the expansion converges (the norm of last term must be smaller than $10^{-20}$). With $\Delta t=0.25\,au$ and $nb_k=10$, the order is about $30$ and the wave packet norm ($N_{1}$) and energy are conserved with accuracies of $10^{-16}$ and $10^{-10}\,au$, respectively. This guarantees that the numerical error of this step is practically zero. Although the size of the basis, $NB$, must be large enough to obtain an accurate wave packet.
\item[(ii)] In the second step, the time-dependent basis parameters $\mathbf{\Lambda}_{t}$ are computed by numerical integration with Gauss-Hermite quadrature grids, and since the number of grid points is larger than the number of basis functions, these parameters can be considered as numerically exact. For our test, we chose $nq_k = nb_k+5$, although a smaller value should be sufficient. The time evolution of these parameters (with $\Delta t=0.25\,au$ and $nb_k=10$) is shown in the following figures (Figs. \ref{fig_QP-2DHarm} and \ref{fig_alpha-2DHarm}) and they show a smooth evolution over time. Furthermore, as expected, all parameters remain constant for coordinates 2 ($k=2)$  along the propagation.
\begin{figure}[hbtp]
\begin{center}
\includegraphics[width=14cm]{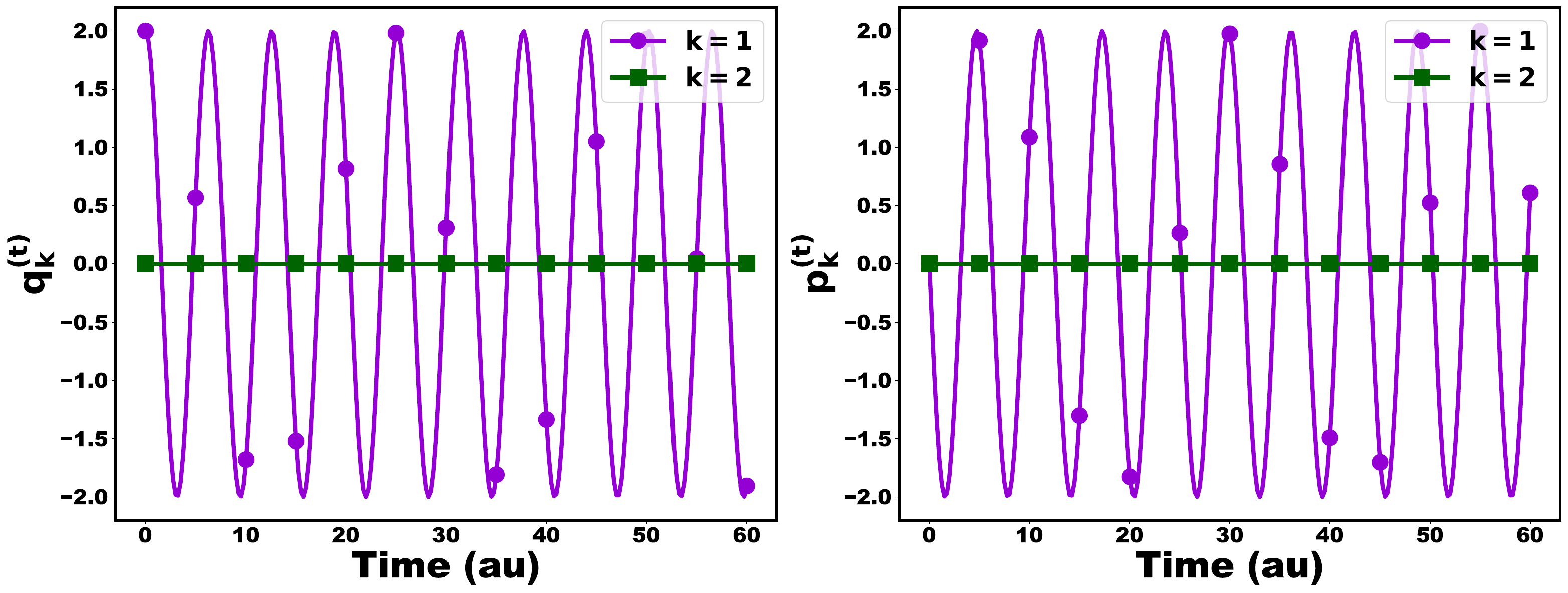}
\caption{\textit{ Time evolution of the Gaussian parameters \( q_k^{(t)} \) (left) and \( p_k^{(t)} \) (right) in the $k^{th}$primitive basis (\(\Delta t = 0.25\,au\), \( nb_k = 10 \)). The expectation values \( q_k^{(t)} \equiv \expval*{\Psi(t) \middle| q_k \middle| \Psi(t) } \) (coordinate) and \( p_k^{(t)} \equiv -\imath \expval*{\Psi(t) \middle| \frac{\partial}{\partial q_k} \middle| \Psi(t) } \) (momentum) are shown for the first (\(k=1\), purple) and second (\(k=2\), green) degrees of freedom. }} \label{fig_QP-2DHarm}
\end{center}
\end{figure}
\begin{figure}[hbtp]
\begin{center}
\includegraphics[width=14cm]{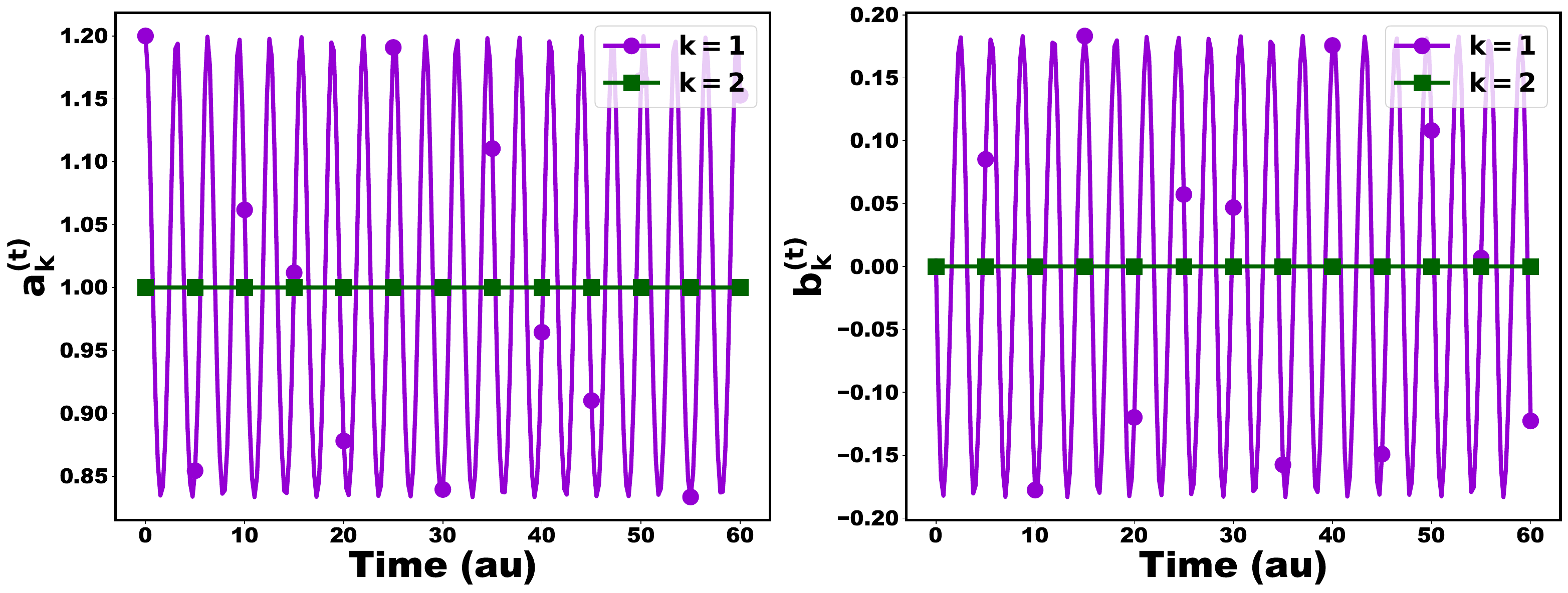}
\caption{\textit{  Time evolution of the complex Gaussian parameter $\alpha_k^{(t)} = a_k^{(t)} + \imath b_k^{(t)}$ components in the $k^{th}$primitive basis. The left panel displays the real part $a_k^{(t)}$ while the right panel shows the imaginary part $b_k^{(t)}$, computed with a time step $\Delta t = 0.25\,au$ and a basis size $nb_k = 10$. The purple and green curves represent the first ($k=1$) and second ($k=2$) degrees of freedom, respectively.}} \label{fig_alpha-2DHarm}
\end{center}
\end{figure}
\item[(iii)] The final step constitutes the most critical part of the procedure. The wave packet $\Psi(\mathbf{Q}, t_2)$ (Eq.~\ref{Eq-psiInter}), computed at time $t_2 = t_1 + \Delta t$ within the basis set parametrized by $\bm{\Lambda}_{t_1}$ (defined at $t_1$), must be subsequently projected onto the time-evolved basis set determined by $\bm{\Lambda}_{t_2}$ (evaluated at $t_2$). The accuracy of this projection depends on the size of the basis set, $NB$ (or the $nb_k$) and the time step, $\Delta t$. The figure(Fig.~\ref{fig_N13-2DHarm}) shows the norm conservation, $\Delta N_{1,3}(t) = \left| N_{1}(t)-N_{3}(t)\right|$ where $N_1(t)$ and $N_3(t)$ denote the norms of the wave packet at time $t$ after the first and third steps of the three-step propagation scheme, respectively. In fact, for a given time step (here $\Delta t=0.25\, au$ ), the projection is only exact if the basis set is large enough. Therefore, one can see that the error decreases to zero as $nb_k$  increases (left panel). Furthermore, for a given basis set size ($nb_k=5$), the error decreases as $\Delta t$ decreases. Indeed, as $\Delta t$ decreases, the wave packet modification between each time step becomes less important, so the new basis set at $t_2$ will be closer to the one at $t_1$. Therefore, the projection will be more accurate.
\begin{figure}[hbtp]
\begin{center}
\includegraphics[width=14cm]{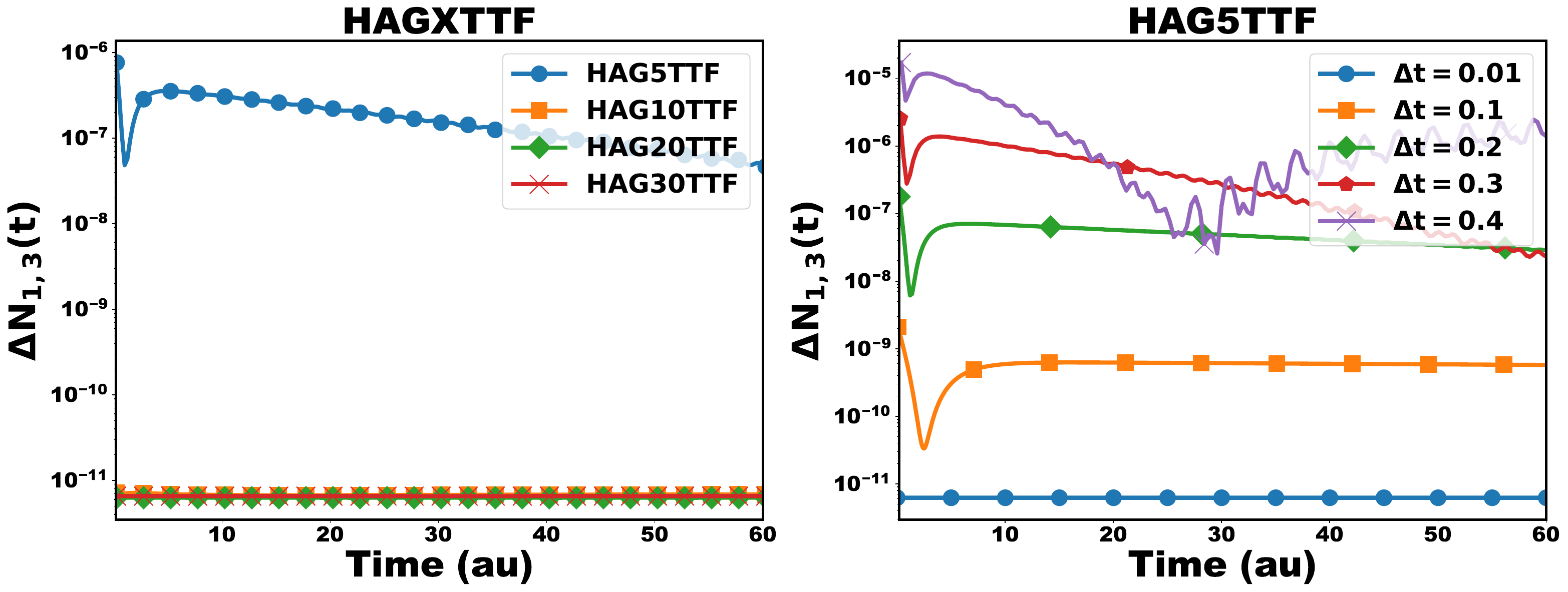}
\caption{\textit{ Time evolution of the wave packet norm deviation $\Delta N_{1,3}(t) \equiv \abs{N_{1}(t)-N_{3}(t)}$ in the three-step propagation scheme. The left panel shows the dependence on basis size $nb_k$ at fixed time step $\Delta t = 0.25\,au$, while the right panel displays the dependence on time step $\Delta t$ for fixed basis size $nb_k = 5$. $N_1(t)$ and $N_3(t)$ denote the norms of the wave packet at time $t$ after the first and third steps of the three-step propagation scheme, respectively.}} \label{fig_N13-2DHarm}
\end{center}
\end{figure}
\end{enumerate}
\subsection{2D-anharmonic Hamiltonian}
\noindent In this section, we use a modified Hénon-Heiles potential with two degrees of freedom. Indeed, in the original Hénon-Heiles potential, the presence of cubic terms leads to holes in the potential. Therefore, to avoid the wave packet delocalization around these holes, we introduced a modified version of the potential (Eq.~ \ref{Eq-HenonHeiles}), in which some $q_k$'s are replaced by $\tanh{\left(q_{k}\right)}$. This modification avoids the holes without changing the first, second and third order derivatives around the main minimum.
{%
\bea \label{Eq-HenonHeiles}
V\left(  \mathbf{Q} \right) =  \frac{mw^2}{2}\sum_{k=1}^{nc}q_{k}^2 
+ \lambda\sum_{k=1}^{nc-1}\left(q_{k}^2 \cdot \tanh{(q_{k+1})} - \frac{1}{3}\tanh{\left(q_{k}\right)}^3\right)
\ena }
The initial wave packet is a Gaussian shifted along the first coordinate (the parameters are given in Table \ref{tab-HOHAGbasis-WP0-2D}) and the propagation time is $\Delta T =60\, au$ with a time step of $dt=0.1\,au$.
The parameters of the potentiel $V\left( \bm{Q}\right) $ of the system are  $m=1.0$, $\omega=1.0$ and $\lambda=0.111803$ \cite{Nest2002}.
\ni For the standard techniques, both coordinates are described using harmonic oscillator (HO) basis sets, while for the Hagedorn scheme, both coordinates are described using Hagedorn basis sets.  
The parameters of the basis sets for both techniques are summarized in Table~\ref{tab-HOHAGbasis-WP0-2D}. For the calculations, we use the number of functions in each primitive basis as \(nb_k = 5, 20, 70\) and for the number of grid points, we add 5 grid points to the number of basis functions so that \(nq_k = nb_k + 5\), \textit{i. e.}, \(nq_k = 10, 25, 75\). This choice of using more grid points than basis functions is made to ensure high accuracy of numerical integration.
\newline 
\begin{table}[H]
	\captionsetup{width=\linewidth}
	\small 
	\caption{\textit{Harmonic Oscillator (HO) and Hagedorn basis set parameter values: $q_k^{(0)}$, $p_k^{(0)}$, and $\alpha_k^{(0)}$ (columns 2--6). Initial Gaussian wave packet parameters (Eq.~\ref{Eq-Psi0}, with $e_0 = 1$): $q_k^0$, $p_k^0$, and $a_k^0$ (columns 7--9)}.}
	\label{tab-HOHAGbasis-WP0-2D}
	\centering
	\begin{tabularx}{\linewidth}{@{}c *{2}{>{\centering\arraybackslash}X} *{3}{>{\centering\arraybackslash}X} *{3}{>{\centering\arraybackslash}X}@{}}
		\toprule
		& \multicolumn{2}{c}{HO basis set} & \multicolumn{3}{c}{Hagedorn basis set} & \multicolumn{3}{c}{Initial wave packet} \\
		\cmidrule(lr){2-3} \cmidrule(lr){4-6} \cmidrule(lr){7-9}
		$k$ & $q_k^{(0)}$ & $a_k^{(0)}$ & $q_k^{(0)}$ & $p_k^{(0)}$ & $\alpha_k^{(0)}$ & $q_k^0$ & $p_k^0$ & $a_k^0$ \\
		\midrule
		\multicolumn{9}{c}{2D} \\
		1     & 2.0  & 1.2  & 2.0  & 0.0  & $(1.2, 0.0)$ & 2.0  & 0.0  & 1.2 \\
		2    & 0.0  & 1.0  & 0.0  & 0.0  & $(1.0, 0.0)$ & 0.0  & 0.0  & 1.0 \\
		\bottomrule
	\end{tabularx}
	\normalsize
\end{table}

\begin{figure}[hbtp]
\begin{center}
\includegraphics[width=14cm]{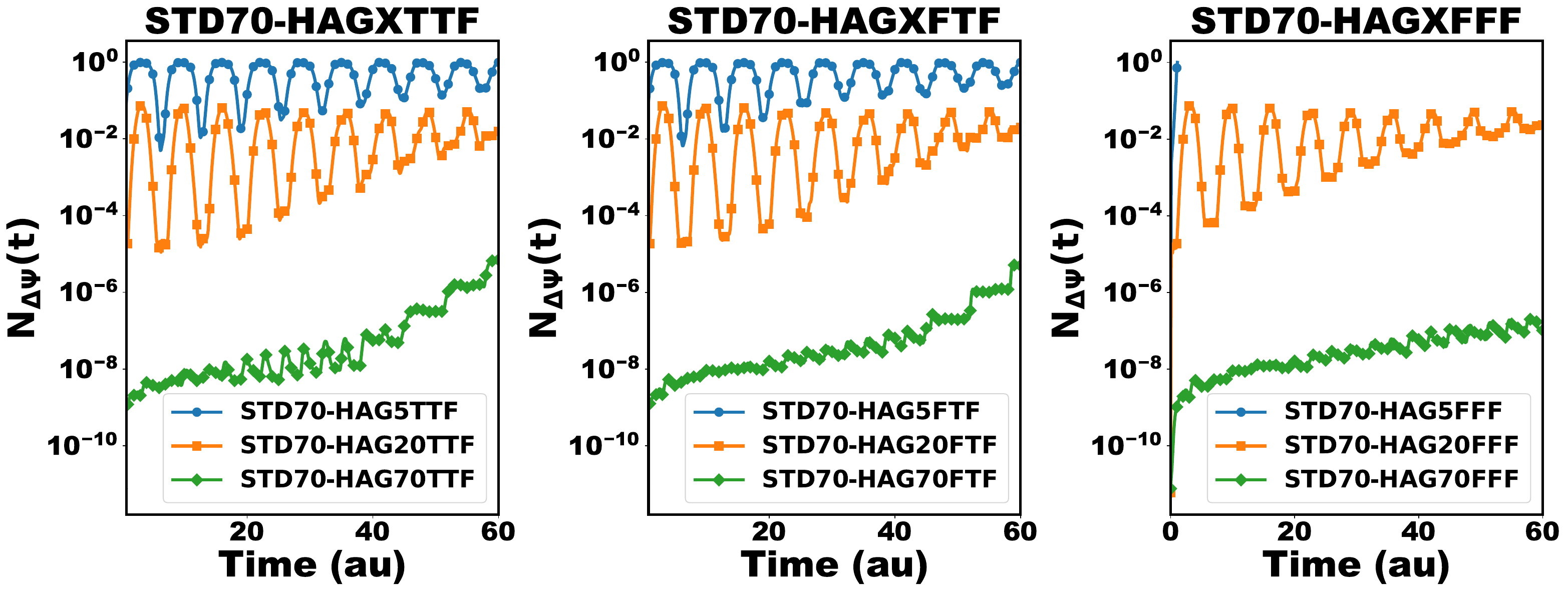}
\caption{\textit{ Time evolution of $N_{\Delta \Psi}(t)$ as a function of $nb$ with $\Delta t=0.1\,au$.  (i) Left panel: with $b^{(t)}_{k}$ and $p^{(t)}_{k}$, (TTF option). (ii) Middle panel: without $b^{(t)}_{k}$ and with $p^{(t)}_{k}$, (FTF option). (iii) Right panel: without $b^{(t)}_{k}$ and without $p^{(t)}_{k}$, (FFF option).}} \label{fig_NDeltaPsi-2DHenonHeiles}
\end{center}
\end{figure}

\begin{figure}[hbtp]
\begin{flushleft}
\includegraphics[width=12cm]{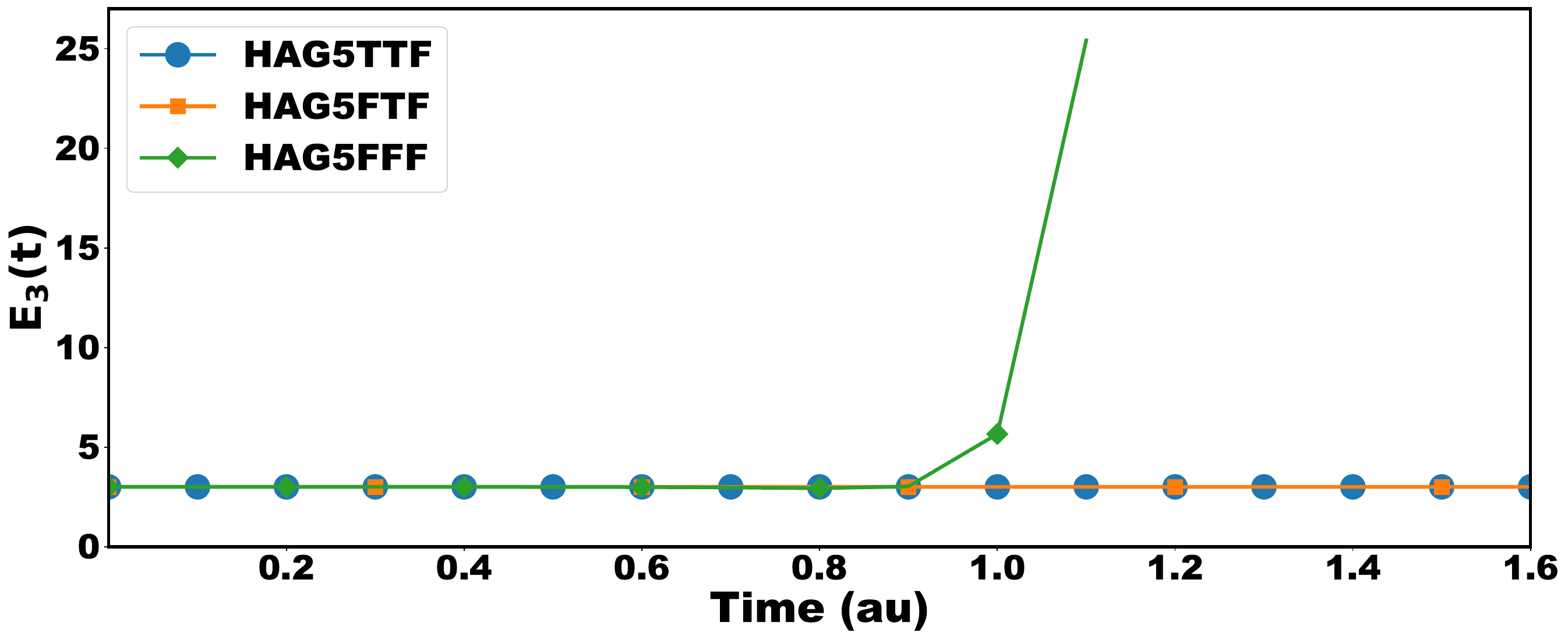}
\caption{\textit{  Energy as a function of time for a propagation with $nb_k=5$:  (i) with $b^{(t)}_{k}$ and $p^{(t)}_{k}$, (TTF option). (ii) without $b^{(t)}_{k}$ and with $p^{(t)}_{k}$, (FTF option). (iii) without $b^{(t)}_{k}$ and without $p^{(t)}_{k}$, (FFF option).}} \label{fig_Energy-2DHenonHeiles}
\end{flushleft}
\end{figure}

\ni With this model, the wave packet does not remain Gaussian, therefore, the accuracy of our procedure is tested with the norm of the difference of two wave packets, $N_{\Delta \Psi}(t)$, namely, the propagation error. The first one, $\ket{\Psi^{std}(t)}$, is obtained using a standard propagation technique (here Taylor expansion), and the second one, $\ket{\Psi^{ProjHag}(t)}$, is obtained after projecting the Hagedorn wave packet onto the basis set used for the standard propagation. This projection is similar to the one in third step of the Hagedorn procedure. More precisely, to obtain $N_{\Delta \Psi}(t)$, we used the coefficients associated with these two wave packets, $\tilde{\mathbf{C}}^t$ (standard propagation) and $\mathbf{C}^t_{ProjHag}$ (projected Hagedorn propagation) and $N_{\Delta \Psi}(t) = \left\lVert \tilde{\mathbf{C}}^t -\mathbf{C}^t_{ProjHag} \right\rVert$.
\ni Before checking the accuracy of the wave packet obtained with the Hagedorn scheme, we need to check that the standard wave packet is converged with respect to the basis set size. Our tests show that with a basis set with 70 basis functions ($nb_k=70$) for each dimension, it is possible to converge the wave packet with an error (with an expression similar to $N_{\Delta \Psi}(t)$) smaller than $10^{-9}$ with respect to a standard propagation with $nb_k=90$.
\ni With this converged wave packet obtained with the standard propagation scheme, we are able to check the accuracy of the Hagedorn scheme (with $\Delta t=0.1\,au$) using several options, but without re-normalization after Hagedorn step 3. In figure \ref{fig_NDeltaPsi-2DHenonHeiles}, we present the propagation error ($N_{\Delta \Psi}(t)$) along time for different options (i) Full Hagedorn scheme, \textit{i. e.} including the variation of $b^{(t)}_{k}$ (the imaginary part of $\alpha^{(t)}_{k}$) and the one of $p^{(t)}_{k}$. This option is labeled (TTF). (ii) A scheme, labeled (FTF), without $b^{(t)}_{k}$ and with the variation of $p^{(t)}_{k}$. (iii) A scheme without $b^{(t)}_{k}$ and without $p^{(t)}_{k}$, labeled (FFF). 
These three options are shown in the figure \ref{fig_NDeltaPsi-2DHenonHeiles} and one can see that for a large basis set, $nb_k \ge 70$, the propagation errors become small for all options ($<5.10^{-6}$ for the TTF and FTF options and  $<10^{-7}$  for the FFF option). This shows the accuracy of our schemes. Furthermore, with this large basis set ($nb_k= 70$) and for all options, the projection error in step 3 (not shown), $\Delta N_{1,3}(t)$, is also very small ($<10^{-11}$). Therefore, the re-normalization after step 3 is not necessary for large basis sets.
\ni For a small value of $nb$ ($nb=5$) and with or without re-normalization after the projection, several points need to be discussed:
(i) It is important to note that the Hagedorn scheme when considering the time variation of $p^{(t)}_{k}$ and with or without the variation of $b^{(t)}_{k}$ (options: (TTF) or (FTF)) is able to propagate up to $60\,au$. The other Hagedorn scheme option (FFF) stops after about $1.\,au$. Indeed, with this option, the wave packet becomes unphysical and the wave packet norm and the total energy are not conserved. The energy conservation error rises by $2.7$ Hartree after a propagation of $1.\,au$ (see Fig. \eqref{fig_Energy-2DHenonHeiles}). Therefore, in the first step of the Hagedorn scheme, the standard propagation with the Taylor series cannot converge (reducing the time step by a factor of 10 does not solve the problem). This problem is due to the projection (step 3 of the Hagedorn scheme), which is less accurate when $p^{(t)}_{k}$ is not included in the Hagedorn scheme. In other words, more basis functions are needed to reach a given accuracy. It is worth noting, that for the (FTF) and (TTF) options, the total energy is conserved up to $2.\times10^{-1}$ and $7.\times 10^{-2}$ Hartree, respectively.
(ii) For the (TTF) or (FTF) options, the propagation error can be very large up to one (Fig. \ref{fig_NDeltaPsi-2DHenonHeiles}), which clearly indicates that the basis set is not large enough without preventing to perform a relatively long propagation. However, we need to check if this will be problematic to compute a spectrum.
\subsection{Comparative Analysis of Energy Spectrum}

\noindent In this section, we also use the modified Hénon-Heiles potential with two degrees of freedom. The basis and initial wave packet parameters are given in the Table \ref{tab-HOHAGbasis-WP0-2D-spectrum}.
\begin{table}[H]
	\captionsetup{width=\linewidth}
	\small 
	\caption{\textit{Harmonic Oscillator (HO) and Hagedorn basis set parameter values: $q_k^{(0)}$, $p_k^{(0)}$, and $\alpha_k^{(0)}$ (columns 2--6). Initial Gaussian wave packet parameters (Eq.~\ref{Eq-Psi0}, with $e_0 = 1$): $q_k^0$, $p_k^0$, and $a_k^0$ (columns 7--9)}}
	\label{tab-HOHAGbasis-WP0-2D-spectrum}
	\centering
	\begin{tabularx}{\linewidth}{@{}c *{2}{>{\centering\arraybackslash}X} *{3}{>{\centering\arraybackslash}X} *{3}{>{\centering\arraybackslash}X}@{}}
		\toprule
		& \multicolumn{2}{c}{HO basis set} & \multicolumn{3}{c}{Hagedorn basis set} & \multicolumn{3}{c}{Initial wave packet} \\
		\cmidrule(lr){2-3} \cmidrule(lr){4-6} \cmidrule(lr){7-9}
		$k$ & $q_k^{(0)}$ & $a_k^{(0)}$ & $q_k^{(0)}$ & $p_k^{(0)}$ & $\alpha_k^{(0)}$ & $q_k^0$ & $p_k^0$ & $a_k^0$ \\
		\midrule
		\multicolumn{9}{c}{2D} \\
		1     & 0.0  & 1.0  & 2.0  & 0.0  & $(1.2, 0.0)$ & 2.0  & 0.0  & 1.2 \\
		2    & 0.0  & 1.0  & 0.0  & 0.0  & $(1.0, 0.0)$ & 0.0  & 0.0  & 1.0 \\
		\bottomrule
	\end{tabularx}
	\normalsize
\end{table}

\begin{figure}[hbtp] 
\makebox[\textwidth]{\includegraphics[width=14cm]{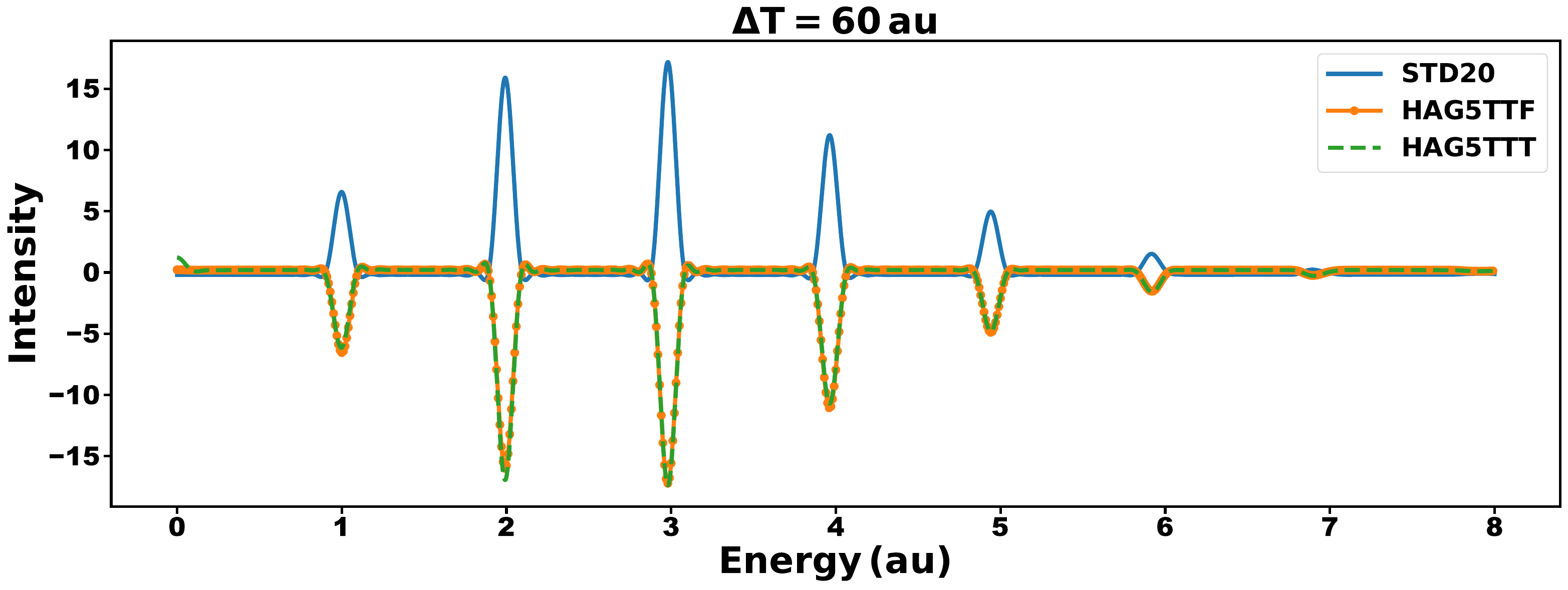}}
\caption{\textit{ Energy spectrum obtained from a propagation of $60\,au$ using: (i) the standard propagation approach with $nb_k = 20$,  labeled STD20 (blue solid line, positive spectrum). (ii) the Hagedorn scheme with $nb_k = 5$, without re-normalization (orange solid line, with orange dot, negative spectrum, labeled HAG5TTF). (iii) the Hagedorn scheme with $nb_k = 5$, with re-normalization (green dashed line, negative spectrum, labeled HAG5TTT).}} 
\label{fig_stdhg5-spectre2D}
\end{figure}

However, the aim is the computation of the spectrum with the Hagedorn scheme and compare it with a reference calculation obtained from a standard calculation. For both approaches, the spectra are computed from the Fourier transform of the autocorrelation function, $a(t)$: 
\be \label{eq:ac}
a(t) = \braket{\Psi(t_0)}{\Psi(t)}
\ee
In addition, and to attenuate the oscillations of each band due to the Gibbs phenomenon, a filter ($\cos^2\left(\frac{\pi(t-t_0)}{t_f-t_0}\right)$, where $t_f$ is the final propagation time) is added to the Fourier transform.
For the standard approach, the  autocorrelation function calculation is just a scalar product between the wave packet coefficients, $\tilde{C}_I$, at $t$ and $t_0$. However, for the Hagedorn scheme, the wave packets at $t$ and $t_0$ do not share the same basis set. Therefore, before performing the scalar product, the wave packet at $t$ developed on the basis set $\mathbf{B}\left( \mathbf{\Lambda}_{t}\right)$, is projected onto the basis set at $t_0$, $\mathbf{B}\left( \mathbf{\Lambda}_{t_0}\right)$ using the technique of the third step of the Hagedorn scheme.
In figure \ref{fig_stdhg5-spectre2D}, three spectra are shown, one from the standard propagation with 20 HO basis functions in each dimension, labeled (STD20) and two from the Hagedorn scheme with 5 basis functions in each dimension with different options. For both propagation approaches, the time step is $0.1\,au$ and the propagation time is $60\,au$.
The first spectrum (blue solid line, positive spectrum, STD20) is our reference spectrum and it shows 6-7 peaks more or less equally spaced by $1.0\,au$. This reflects the small anharmonicity of the potential. Furthermore, the width of each band is too large to distinguish its components. 
The second (labeled HAG5TTF) and third  (labeled HAG5TTT) spectra are obtained using the Hagedorn scheme with $b^{(t)}_{k}$ and $p^{(t)}_{k}$ and without or with re-normalization, respectively. The figure  \ref{fig_stdhg5-spectre2D} shows that the spectra obtained using both Hagedorn approaches accurately reproduce the spectrum computed with the help of the standard method.
%

\begin{figure}[hbtp]
\makebox[\textwidth]{\includegraphics[width=14cm]{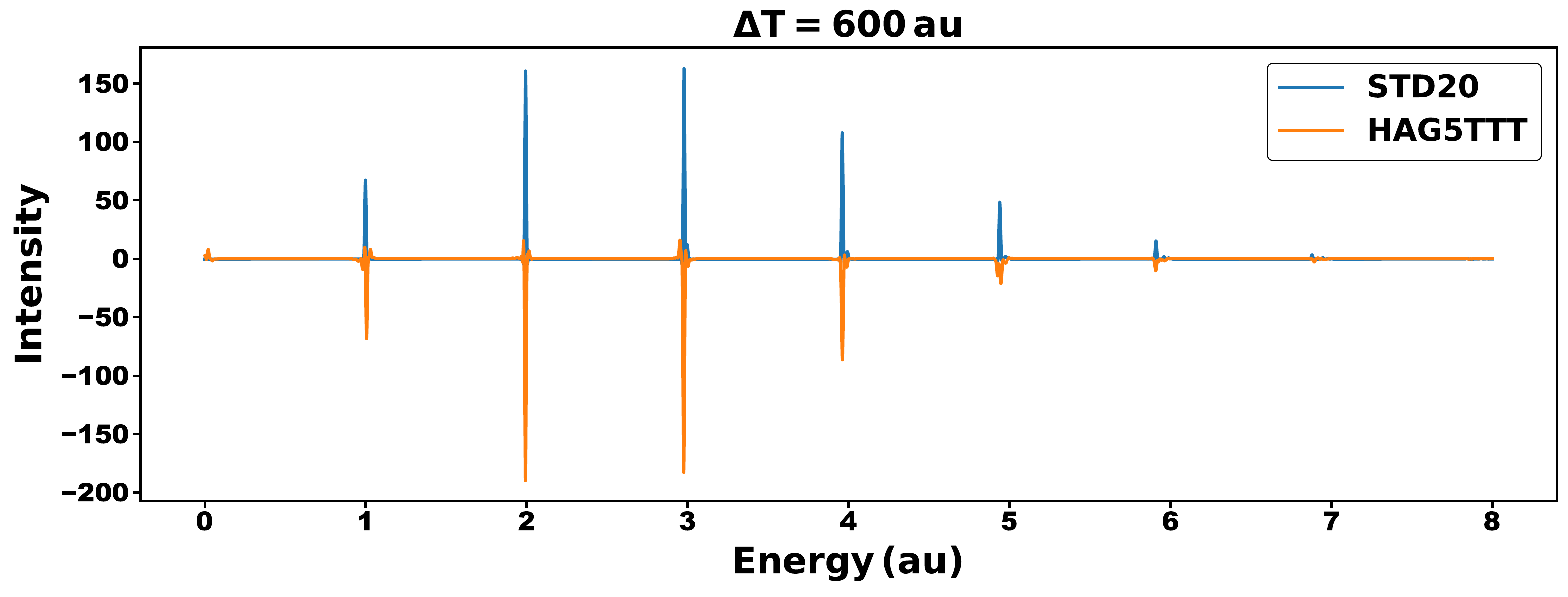}}
\caption{\textit{ Energy spectrum obtained from a propagation of $600\,au$ using: (i) the standard propagation approach with $nb_k = 20$,  labeled STD20 (blue solid line). (ii) the Hagedorn scheme with $nb_k = 5$ and with re-normalization (orange solid line,  labeled HAG5TTT).}} 
\label{fig_stdhg5-long-spectre2D}
\end{figure}

\begin{figure}[hbtp]
\centering
\includegraphics[width=14cm]{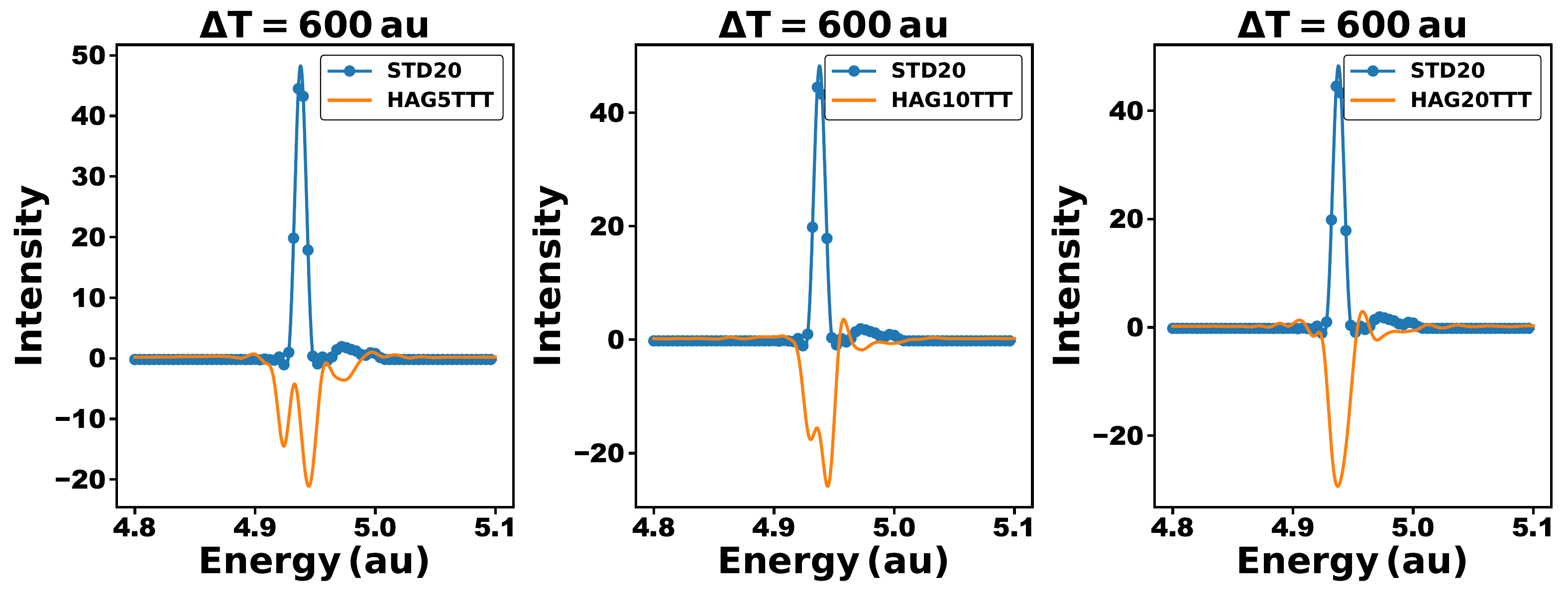}
\caption{\textit{ Energy spectrum around $5\,au$ obtained from a propagation of $600\,au$ using: (i) the standard propagation approach with $nb_k = 20$,  labeled STD20 (blue solid line,with blue dot). (ii) the Hagedorn scheme with re-normalization (orange solid line). Left panel with $nb_k = 5$ (HAG5TTT), middle panel with $nb_k = 10$ (HAG10TTT) and right panel with $nb_k = 20$ (HAG20TTT).}}
\label{fig_stdhg51020-spectre2D}
\end{figure}

\begin{figure*}[hbtp]
\makebox[\textwidth]{\includegraphics[width=15cm]{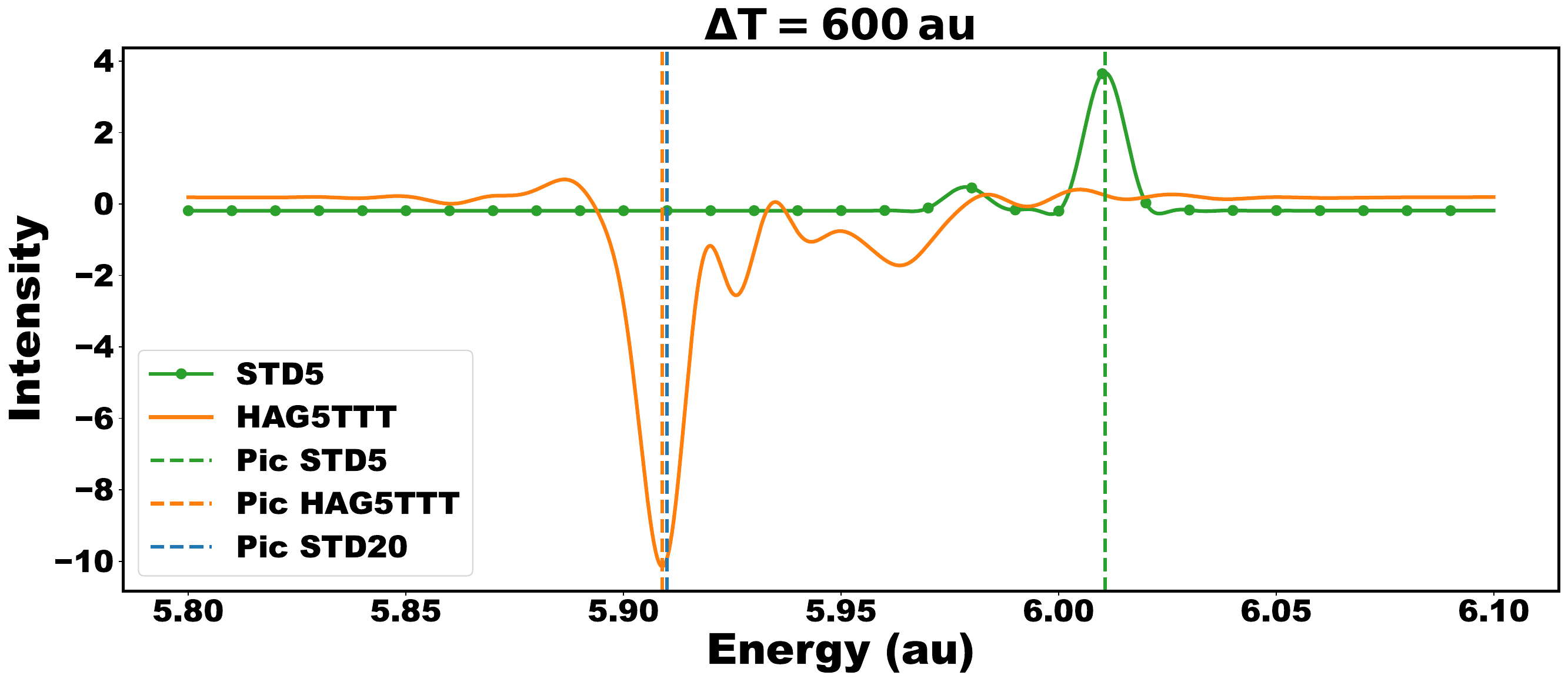}}
\caption{\textit{ Energy spectrum around $6.0 \,au$ obtained from a propagation of $600 \,au$ using: (i) the standard propagation approach with $nb_k = 5$, labeled STD5 (green solid line, with green dot). (ii) the Hagedorn scheme with re-normalization (orange solid line) with $nb_k = 5$ (HAG5TTT). In addition, vertical lines show the main peak positions of STD5 (green dashed line), STD20 (blue dashed line) and HAG5TTT (orange dashed line), respectively.}} 
\label{fig_std520hg20-spectre2D}
\end{figure*}

Finally, to further analyze the Hagedorn scheme, we perform propagations with a larger final propagation time of $600\,au$ instead of $60\,au$ for the previous calculations. Since, we have obtained a good spectrum without  re-normalization for short propagation of $60\,au$, we have tried the long propagation under the same conditions (HAG5TTF). Unfortunately, the propagation have stopped before reaching the final time (the total energy has increased by $0.6$ Hartree). Therefore, we have performed the propagation with re-normalization (HAG5TTT). The corresponding spectrum (orange solid line, labeled HAG5TTT) compares very well with the STD20 one (see Fig. \ref{fig_stdhg5-long-spectre2D}). 
\ni In addition, the figure \ref{fig_stdhg51020-spectre2D} shows the spectrums around $5\,au$ for the standard approach and for the Hagedorn scheme with several basis set sizes. The spectrum obtained with the standard approach shows a main peak and a small side band at higher energy due to the five components around this energy. However, the first Hagedorn spectrum (with $nb_k=5$, HAG5TTT, left panel) shows a peak on the left that is not present in the reference spectrum. The other two peak positions are well reproduced, although the intensity of the central peak is too small. When the basis set size increases, this additional peak disappears (Fig. \ref{fig_stdhg51020-spectre2D}, central and right panels).
\ni The last figure (Fig. \ref{fig_std520hg20-spectre2D}) of this section shows spectra around 6. au obtained with only 5 basis functions in each dimension for both the standard (STD5) and the Hagedorn (HAG5TTT) approaches. The position of the Hagedorn main peak (marked with an orange vertical line) is in excellent agreement with that of the standard approach obtained with $nb_k=20$ (blue vertical line). However, the main peak position of the STD5 calculation is strongly overestimated. This is to be expected, since the basis set is too small to reproduce the 6 components of the energy levels around $6\,au$. 
%
\section{Application on the 6D Hénon-Heiles model}  \label{Sec-Appli}

 In the previous section, the propagation was done with models with two degrees of freedom. In this section, we compare the Hagedorn scheme with the standard approach using the modified Hénon-Heiles model (Eq. \ref{Eq-HenonHeiles}) with six degrees of freedom \cite{Nest2002}.
For these calculations, the propagation time is $60\,au$ and the time step is $0.1\,au$. The basis set and the initial wave packet parameters are given in the Table \ref{tab-HOHAGbasis-WP0-6D} and differ slightly from the 2D ones.

\begin{table}[H]
	\captionsetup{width=\linewidth}
	\small 
	\caption{\textit{Harmonic Oscillator (HO) and Hagedorn basis set parameter values: $q_k^{(0)}$, $p_k^{(0)}$, and $\alpha_k^{(0)}$ (columns 2--6). Initial Gaussian wave packet parameters (Eq.~\ref{Eq-Psi0}, with $e_0 = 1$): $q_k^0$, $p_k^0$, and $a_k^0$ (columns 7--9).}}
	\label{tab-HOHAGbasis-WP0-6D}
	\centering
	\begin{tabularx}{\linewidth}{@{}c *{2}{>{\centering\arraybackslash}X} *{3}{>{\centering\arraybackslash}X} *{3}{>{\centering\arraybackslash}X}@{}}
		\toprule
		& \multicolumn{2}{c}{HO basis set} & \multicolumn{3}{c}{Hagedorn basis set} & \multicolumn{3}{c}{Initial wave packet} \\
		\cmidrule(lr){2-3} \cmidrule(lr){4-6} \cmidrule(lr){7-9}
		$k$ & $q_k^{(0)}$ & $a_k^{(0)}$ & $q_k^{(0)}$ & $p_k^{(0)}$ & $\alpha_k^{(0)}$ & $q_k^0$ & $p_k^0$ & $a_k^0$ \\
		\midrule
		\multicolumn{9}{c}{6D} \\
		1     & 0.0  & 1.0  & 2.0  & 0.0  & $(1.0, 0.0)$ & 2.0  & 0.0  & 1.0 \\
		2-6& 0.0  & 1.0  & 0.0  & 0.0  & $(1.0, 0.0)$ & 0.0  & 0.0  & 1.0 \\
		\bottomrule
	\end{tabularx}
	\normalsize
\end{table}

\begin{figure}[hbtp]
\centering
\makebox[\textwidth]{\includegraphics[width=15cm]{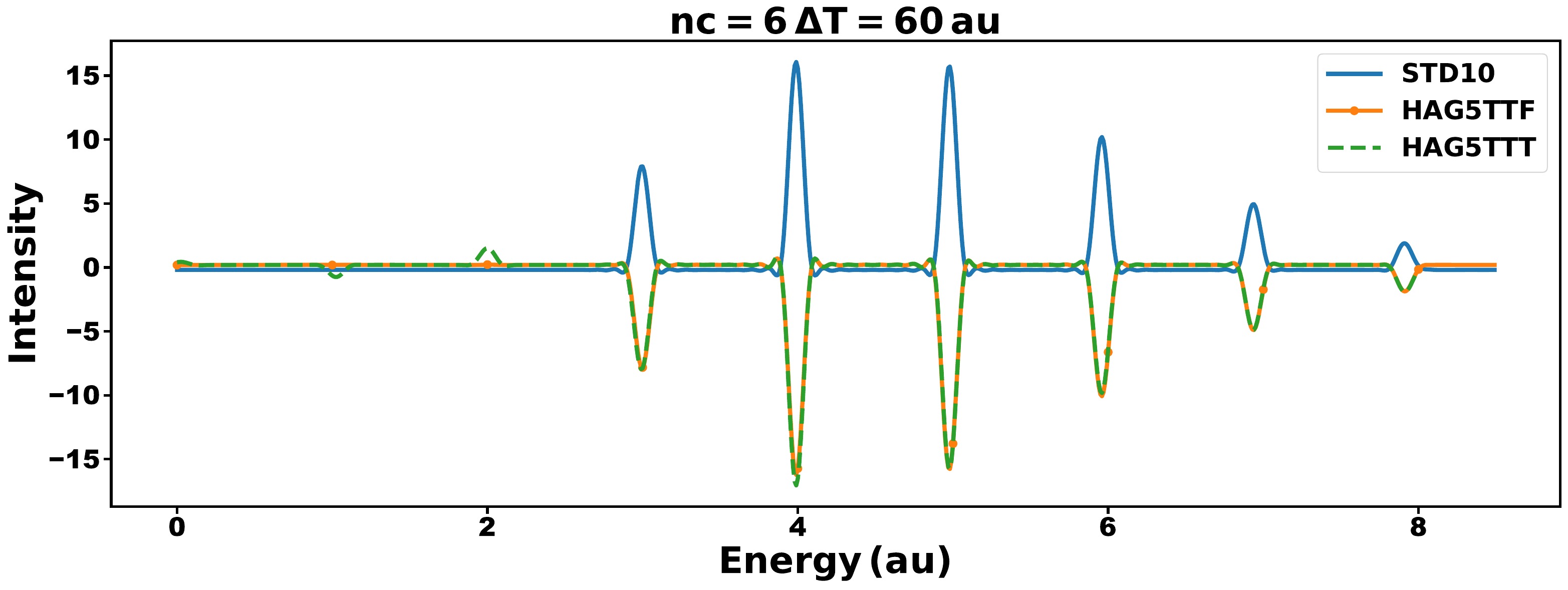}}
\caption{\textit{ Energy spectrum obtained from a propagation of $60\,au$ using: (i) the standard propagation approach with $nb_k = 10$,  labeled STD10 (blue solid line, positive spectrum). (ii) the Hagedorn scheme with $nb_k = 5$, without re-normalization (orange solid line, with orange dot, negative spectrum, labeled HAG5TTF). (iii) the Hagedorn scheme with $nb_k = 5$, with re-normalization (green dashed line, negative spectrum, labeled HAG5TTT).}}
\label{fig_std10hg5-spectre6D}
\end{figure}

\begin{figure}[hbtp]
\centering
\makebox[\textwidth]{\includegraphics[width=15cm]{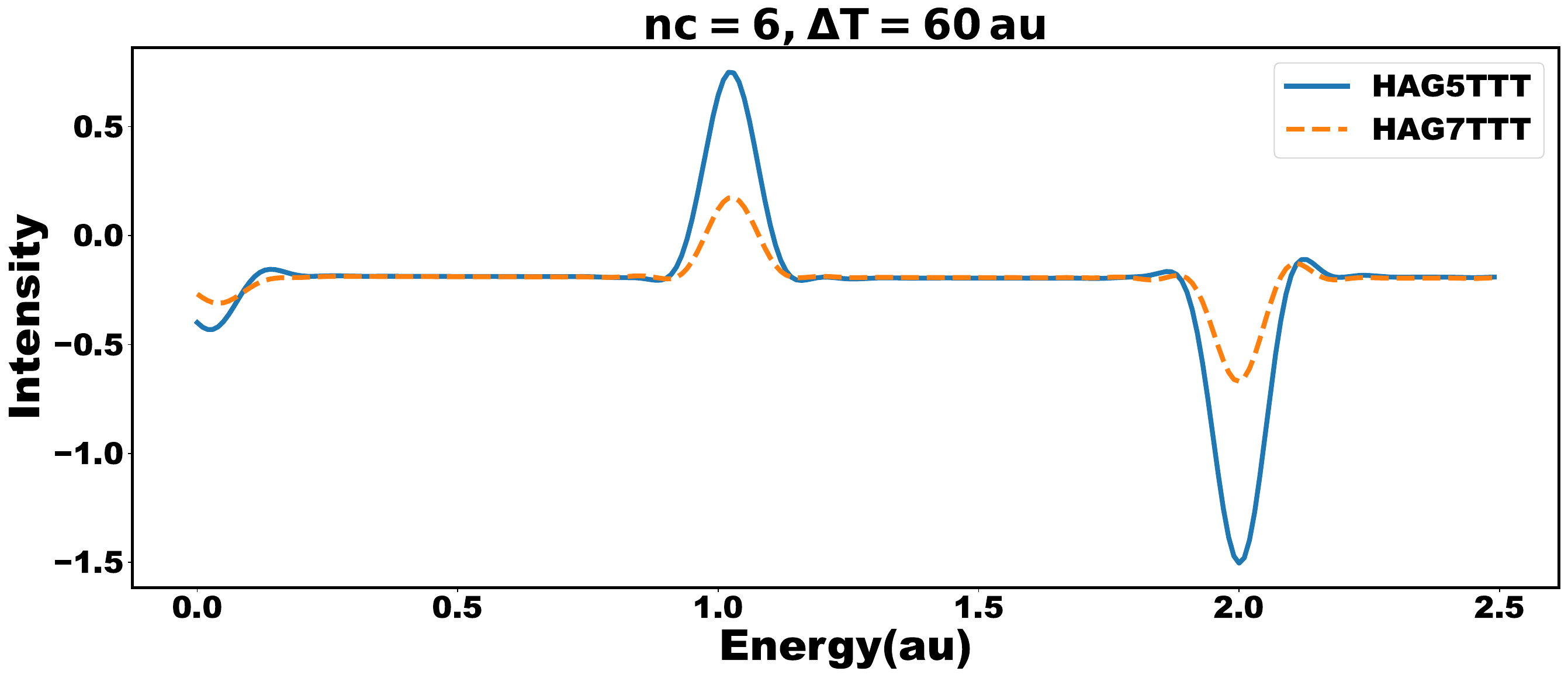}}
\caption{\textit{ Energy spectrum obtained from a propagation of $60 \,au$  for energies below $2.5 \,au$ with the Hagedorn scheme with re-normalization and with $nb_k = 5$ (blue solid line, labeled HAG5TTT) or with $nb_k = 7$ (orange dashed line, labeled HAG7TTT).}}
\label{nc6hag57-spectrum-zoom}
\end{figure}

The figure \ref{fig_std10hg5-spectre6D} shows the spectrum (blue solid line, SDT10) for the standard propagation with 10 basis functions in each dimension, as well as, the spectra for the Hagedorn scheme with  $nb_k = 5$ (green dashed line, HAG5TTT and orange solid line, with orange dot, HAGTTF).
As with the 2D-model (Fig. \ref{fig_stdhg5-spectre2D}), and with this short propagation time, the agreement is almost perfect. However, two non-physical peaks  appear at an energy below the first physical level (around $3.\,au$) in the spectra obtained with re-normalization (HAG5TTT).  It is important to note that these two peaks are absent from the spectra obtained without re-normalization (HAG5TTF). We believe, these two peaks are due to the deformation of the autocorrelation function resulting from the wave packet re-normalization. To confirm this deformation, we have computed the error of the Hagedorn autocorrelation function with respect to the standard autocorrelation function as follows: 
\be \label{eq:ac_err}
Err = \frac{\displaystyle \int_0^{t_f} \left| a^{\mathrm{STD}}(t) - a^{\mathrm{HAG}}(t) \right| \, dt}{t_f},
\ee
where, $a^{\mathrm{STD}}(t)$ and $a^{\mathrm{HAG}}(t)$ denote the auto-correlation functions obtained from the standard and Hagedorn calculations, respectively. 
The errors associated to the Hagerdorn calculation with re-normalization (HAG5TTT) is about twice larger than the one without re-normalization (HAG5TTF). 
Furthermore, the intensities of these non-physical peaks are reduced when the basis set sizes increase. Figure \ref{nc6hag57-spectrum-zoom} illustrates this reduction as the basis set sizes increase from $nb_k = 5$ to $nb_k = 7$. This feature is expected, as a larger basis set reduces the loss of wavepacket norm after the third step (see this effect for the Harmonic model in Fig \ref{fig_N13-2DHarm}).

\begin{figure}[hbtp]
\centering
\makebox[\textwidth]{\includegraphics[width=15cm]{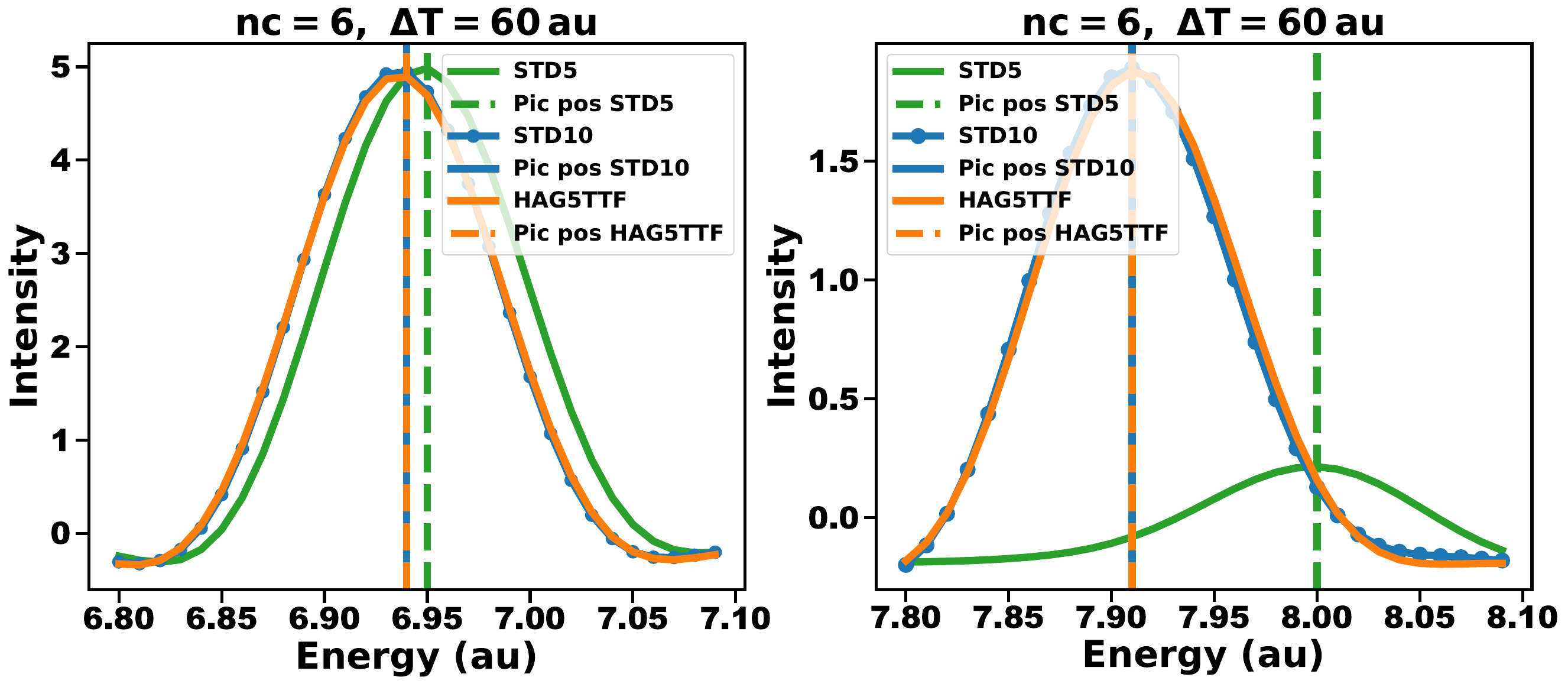}}
\caption{\textit{ Energy spectra around $7.0 \,au$ (left panel) and around $8.0 \,au$ (right panel) obtained from a propagation of $60 \,au$ with: (i) the standard propagation approach with $nb_k = 5$, labeled STD5 (green solid line), (ii) the standard propagation approach with $nb_k = 10$, labeled STD10 (blue solid line). (iii) the Hagedorn scheme with re-normalization (orange solid line) with $nb_k = 5$ (HAG5TTF). In addition, vertical lines show the main peak positions of the STD5 (green dashed line), the STD10 (blue dashed line) and the HAG5TTF (orange dashed line), respectively.}}
\label{nc6std510hg5ttt-spectrum}
\end{figure}

\ni  Finally, the last figure (Fig. \ref{nc6std510hg5ttt-spectrum}) shows spectra around $7.$ and $8.$ Hartree obtained with only 5 basis functions in each dimension for both the standard (STD5) and the Hagedorn (HAG5TTF) approaches and as well as the reference spectrum (STD10). As for the 2D-model, the position of the Hagedorn main peak (marked with an orange vertical line) is in excellent agreement with that of the standard approach obtained with $nb_k=10$ (blue vertical line). However, the main peak positions of the STD5 calculation are slightly overestimated at $7.\,au$ and strongly overestimated at $8.\,au$. 
For this 6D-anharmonic model, the Hagedorn scheme allows the use of 1D-basis sets, which are approximately two times smaller than those used in the standard approach. Therefore, the 6D-Hagedorn basis set is approximately 64 times smaller than the one used in the standard scheme. However, in the Hagedorn scheme, the potential representation must to be recomputed at each time step because the basis set (and thus the grid) changes. Nevertheless, the total Hagedorn propagation is approximately three times faster than the standard propagation. Regarding the accuracy of the peak positions obtained from the HAG5TTF calculation compared to the STD10 calculation, the largest error is approximately $3\times 10^{-3}\,au$ (a relative error of $0.03\,\%$) for the peak around $8.\,au$.
\section{Conclusion and perspectives}  \label{Sec-Conclu}
\noindent In this study, we present a numerical implementation of the Hagedorn wave packet with the possibility to use a combination of time-dependent basis sets (Hagedorn ones) and time-independent basis sets (standard ones) with one or several diabatic electronic states. Some remarks are in order with respect to the actual numerical implementation:
\begin{itemize}
\item The propagation is performed in three steps: (i) standard propagation (ii) parameter computation of the time-dependent basis set, assuming the wave packet remains Gaussian (iii) projection of the wave packet onto the new basis set. On the one hand, this strategy does not rely on the time-dependent variational principle, so we do not need to compute an inverse of the overlap matrix, which may be singular. On the other hand, this scheme is not optimal because the time-dependent parameters are computed assuming that the wave packet remains Gaussian, which is not the case for realistic systems. Therefore, we are planning to implement the time-dependent variational principle \cite{mclachlan1964time} to propagate the time-dependent parameters.
\item Since we are using grid, we are not formally restricted in the choice of the Hamiltonian form. Indeed, we can use simple analytical models (as we did in the present study), but also more complex ones for which the potentials do not have simple expressions. Furthermore, we can easily extend our implementation so that complex curvilinear coordinates can be used with analytical kinectic energy operator (KEO) expressions \cite{Wilson1955a, Meyer1968, Chapuisat1991, Chapuisat1992a, Wang2000a, Gatti2001a, Ndong2012, Ndong2013} or with a KEO obtained numerically \cite{Meyer1979, Harthcock1982, Senent1995, Luckhaus2000, Lauvergnat2002, Matyus2009, Marsili2022}. 
Indeed, the first step of our propagation scheme can be used with any Hamiltonian form (as it is done in discrete variable representation or pseudo-spectral approaches \cite{gottlieb1977numerical, light2000discrete, friesner1986solution, szalay2012variational, yu2005coherent, gatti1999fully}) and the other two steps are independent of the Hamiltonian.
\end{itemize}
\noindent This implementation has been tested on several models. In particular, it was able to reproduce the Heller scheme for the quadratic model. This test was important to check the numerical stability of our three-step scheme. The tests on anharmonic systems (2D and 6D) show that it was able to reproduce a standard propagation scheme with high accuracy. Furthermore, the spectra, computed from the Fourier transform of the autocorrelation function, are also well reproduced, but with a smaller basis set than the one required for the standard propagation scheme. 
In addition, with the direct product expansion of the wave packet, the number of operations required to compute the Hamiltonian action on a wave packet grows as $nc \cdot {n_k}^{nc+1}$\cite{Light2000}. Therefore and assuming a 1D-Hagedorn basis set two times smaller than the standard one, we can expect a speedup of approximately of $2^{nc+1}$ for the Hagedorn scheme with respect of the standard one. However, this has to be reduced since the potential must be recalculated with each time step in the Hagedorn scheme.

\noindent As mentioned in section \ref{Sec-Meth}, our implementation allows the study of models with several electronic states. To show this feature, we test our implementation with the 2D-model of retinal photoisomerization developed by Susanne Hahn and Gerhard Stock \cite{hahn2000quantum} (see the appendix \ref{2D-retinal}). This model has played a key role in understanding ultrafast photochemical processes in biological systems and was later extended to a 2+23D model that includes additional vibrational modes, providing a more comprehensive description of the system \cite{hahn2000femtosecond}. Over the years, this model has been extensively studied and refined, contributing significantly to the field of photoinduced molecular dynamics \cite{Pereira2023,johnson2017primary,seidner1994microscopic,domcke1997theory,balzer2003mechanism}.

\noindent This model is described by a $2 \times 2$ Hamiltonian in the diabatic basis with two coordinates, $\Phi$ associated with the torsion (it describes the cis-trans isomerization) and $Q$ associated with the active vibronic coupling mode. For this system, we use a Fourier basis set for the $\Phi$ coordinate (with $nb_{\Phi}=nq_{\Phi}=256$) for both the standard and Hagedorn propagation approaches. For the coordinate $Q$, we use the HO basis set and the Hagedorn one for the standard and Hagedorn propagation schemes, respectively. The size of the basis set is denoted by $nb_Q$ and the corresponding grid size, $nq_Q$, is $nb_Q+5$. Initially, the system is assumed to have already absorbed light from the cis isomer. Therefore, the initial wave packet is expressed as a 2D-Gaussian on the second diabatic electronic state ($e_0=2$). The basis set and initial wave packet parameters are given in the appendix \ref{2D-retinal}.\\

\begin{figure}[hbtp]
\begin{center}
\makebox[\textwidth]{ \includegraphics[width=16cm]{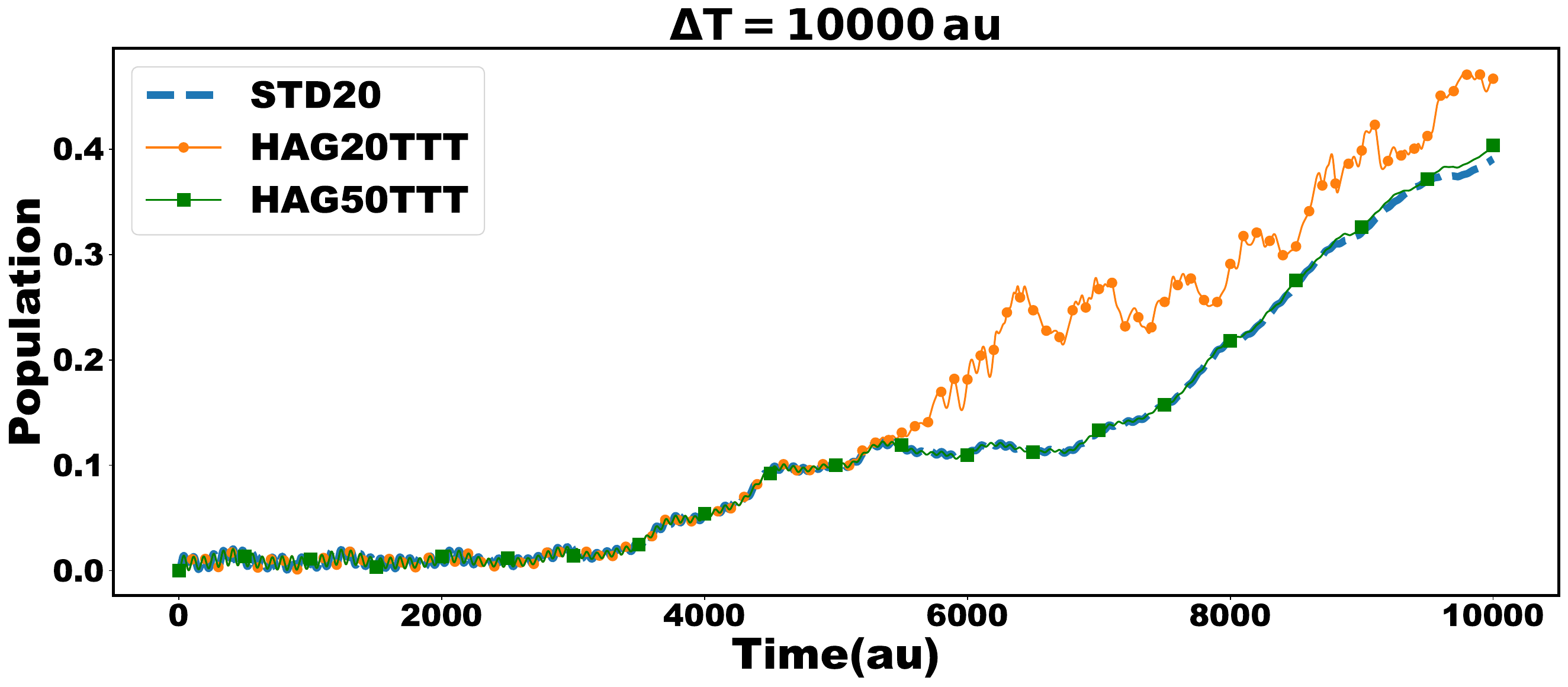}}
\end{center}
\caption{\textit{Time evolution of the ground state population during retinal isomerization obtained from two computational approaches: (i) the standard method (STD20, blue dashed line) and (ii) the Hagedorn approach with two variants (HAG20TTT, orange solid line with solid dot; HAG50TTT, green solid line with squares).}}
\label{fig_nc2rethag}
\end{figure}

\noindent For the comparison between the standard and the Hagedorn propagation scheme, we evaluate the populations, $p_e$, along time for each diabatic state $e$ ($p_e= \braket{\psi^{(e)}(t)}{\psi^{(e)}(t)}$).
For a propagation of $10000.\,au$ (with a time step of $1.\,au$), the numerical convergence is reached with $nb_Q = 20$ and $nb_Q = 50$ for the standard and Hagedorn propagation schemes, respectively (see the first diabatic state population in the Fig. \ref{fig_nc2rethag}). The fact that more basis functions are needed for the Hagedorn scheme seems surprising. However, since we have a common basis set for both diabatic states (equivalent to the single-set formalism in MCTDH), the Hagedorn basis parameters are better adapted to the state with a large population (here the second one). To overcome this difficulty, we are planing to describe the wave packet with different basis sets for the different electronic states (equivalent to the multi-set formalism in MCTDH). Furthermore, the spawning technique, which was originally developed for a wave packet described by a sum of multidimensional Gaussian wave function \cite{Ben-Nun1998,curchod2018ab}, can be adapted to the Hagedorn wave packet approach, as shown by Gradinaru et al. \cite{Gradinaru2024}.\\

 In the present study, the initial wave packet is described as a product of 1D-Gaussian along the coordinates. Although, this approximation is widely used (in particular for non-adiabatic dynamics \cite{Cigrang2025}), a relaxation approach can easily be implemented to obtain the  vibrational or vibronic ground state with the Hagedorn approach. Furthermore, the photoinduced processes can be modelized through a propagation: 
 	(i) for which the initial wave packet is obtained by the action of the transition dipole moment operator on the ground vibronic state. 
 	(ii) with a time-dependent Hamiltonian\cite{lauvergnat2007simple} including  the time-dependent electric field explicitly with a semiclassical field/matter description. As mentioned in the introduction, the aim of this study is the development of a Hagedorn wave packet technique and therefore at a temperature of 0 K. Developing such Hagedorn technique to solve Liouville–von Neumann equation is certainly doable but beyond the scope of this study. Although the Monte Carlo Wave-Function (MCWF)\cite{Dalibard1992} or the Variational Wave Packet (VWP)\cite{Gerdts1997} approaches have been developed to efficiently solve the Liouville–von Neumann equation.
 
\quad \newline

Furthermore, the actual implementation relies on a direct-product of the basis sets and grids. Therefore, it is limited to small system sizes since the total number of basis functions, $NB$, or grid points, $NQ$, grows exponentially with the number of degrees of freedom, $nc$, although this is somehow attenuated since a small basis set size could be used when spectra are computed.
To overcome this curse of dimensionality, a basis function selection scheme has already been implemented for Hagedorn wave packets using algebraic approaches \cite{vanivcek2024hagedorn}. However, in our actual implementation, a standard approach with a direct-product multidimensional basis set is used. A pruned basis set or a basis set with a selection of basis functions is insufficient since a multidimensional grid is required. Therefore, a Smolyak scheme \cite{Smolyak1963a} combined with standard propagation techniques\cite{Gradinaru2008,lasser2020computing,Pereira2023,Xie-Chen2024} could be well-adapted, as it avoids exponential scaling for both the basis set and the grid.
The Smolyak technique was first applied to vibrational spectroscopy by Avila and Carrington, followed by several other groups \cite{Avila2009,Avila2017a,Lauvergnat2013,Avila2019,Chen2022,Lauvergnat2023}. Additionally, the Smolyak scheme has been applied to wave packet propagation, both with and without non-adiabatic couplings \cite{Gradinaru2008,lasser2020computing,Pereira2023,Xie-Chen2024}. Therefore, we are also planning to combine the Smolyak scheme with the Hagedorn propagation technique.
With this improvement, we should be able to study larger systems, such as spin-boson models, although with a finite bath size. The Smolyak approach has already been applied to system/bath models. \cite{Chen2022,Pereira2023}.
\section*{Acknowledgments}
K. S., R. I. and K. M. R. A. thank the Institut de Chimie Physique (ICP) of the Université Paris Saclay and the CNRS, France, for computing facilities.
\noindent  R. I. and K. M. R. A. gratefully acknowledge the ICP for hosting our research stays and also the ThéoSim group for their scientific support and hospitality.
\noindent This work was supported through mobility grants from the Service de Coopération et d'Action Culturelle (SCAC) of the French Embassy in Togo. We additionally thank Campus France for their administrative assistance during our stays in France.

\section*{Disclosure statement}

No potential conflict of interest was reported by the author(s).

\section*{Data availability statement}

Data sharing is not applicable to this article as no new data were created or analyzed in this study.

\appendix

\section{Appendix: Projection} \label{WPProject}
\noindent Once the new basis is constructed, we can proceed with the projection. From Eqs \ref{Eq-psiInter},\ref{Eq-psifinal},\ref{Eq-projectcoef}, we recall that, the $C^{t_2}_{I}$ are given by:
{
\begin{equation}
C^{t_2}_{J} = \braket*{B_{J}\left(\bm{\Lambda}_{t_2}\right) }{\Psi(t_2)}=\sum_{I=1}^{NB} \tilde{C}^{t_2}_{I} \braket*{B_{J}\left(\bm{\Lambda}_{t_2}\right) }{B_{I}\left(\bm{\Lambda}_{t_1}\right) },
\end{equation}}
where, $\tilde{C}^{t_2}_{I}$ and $C^{t_2}_{J}$ are the wave packet coefficients at time $t_2$ associated, respectively, to the $I^{th}$ basis function at time $t_1$ and  $J^{th}$ basis function at time $t_2$.  
This equation shows that the projection is performed on a multidimensional basis. The key elements are the overlaps between the basis functions at different times, $t_1$ and $t_2$. However, numerically, it can be achieved sequentially using partial projections (one coordinate after another). To perform the sequential transformations, the coefficients $\tilde{C}^{t_2}_{I}$ and $C^{t_2}_{I}$ notations need to be re-expressed as :
\begin{eqnarray}
\begin{aligned}
    \tilde{C}^{t_2}_{I} &\equiv C_{i_1, \cdots i_k, \cdots, i_{nc}}^{\bm{b}^1_{t_1} \cdots \bm{b}^k_{t_1}, \cdots, \bm{b}^{nc}_{t_1}}(t_2), \\
    C^{t_2}_{I}            &\equiv C_{i_1, \cdots i_k, \cdots, i_{nc}}^{\bm{b}^1_{t_2} \cdots \bm{b}^k_{t_2}, \cdots, \bm{b}^{nc}_{t_2}}(t_2).
\end{aligned}
\end{eqnarray}
This notation enables to show that the coefficients come from primitive basis sets at different times $t_1$ or $t_2$ .Furthermore, we recall that:
{ 
\begin{eqnarray}
\begin{aligned}
B_{I}\left(\bm{Q},\bm{\Lambda}_{t_1}\right) &\equiv b^1_{i_1}\left( q_1,t_1\right)  \cdots  b^k_{i_k}\left( q_k,t_1\right)  \cdots  b^{nc}_{i_{nc}}\left( q_{nc},t_1\right),  \\
B_{I}\left(\bm{Q},\bm{\Lambda}_{t_2}\right) &\equiv  b^1_{i_1}\left( q_1,t_2\right)  \cdots  b^k_{i_k}\left( q_k,t_2\right)   \cdots  b^{nc}_{i_{nc}}\left( q_{nc},t_2\right) .
\end{aligned}
\end{eqnarray}}
\noindent The functions $b^k_{i_k}\left( q_k,t_1\right)  \equiv b^k_{i_k}\left( q_k,\lambda_k^{(t_1)}\right)  $ and $b^k_{i_k}\left( q_k,t_2\right)  \equiv b^k_{i_k}\left( q_k,\lambda_k^{(t_2)}\right) $ represent the $i_k^{\text{th}}$ function of the primitive basis $k$ at time $t_1$ and time $t_2$, respectively. 
\noindent With these notations, the only difference between the $\tilde{C}^{t_2}_{I}$ and $C^{t_2}_{I}$ or between the $B_{I}\left(\bm{Q},\bm{\Lambda}_{t_1}\right)$ and $B_{I}\left(\bm{Q},\bm{\Lambda}_{t_2}\right)$ are the time indices ($t_1$ or $t_2$) of the $\lambda_k^{t}$ time dependent basis parameters ($k$). \\
The partial projection for the coordinate $k$ reads: 
\begin{eqnarray}
&& C_{j_1, \cdots j_k, i_{k+1} \cdots, i_{nc}}^{\bm{b}^1_{t_2} \cdots \bm{b}^k_{t_2}, \bm{b}^{k+1}_{t_1}, \cdots, \bm{b}^{nc}_{t_1}}\left( t_2\right)  = \sum_{i_k}
\braket{\bm{b}^k_{j_k}\left(t_2\right) }{\bm{b}^k_{i_k}\left(t_1\right) } \cr
& &\qquad \quad \times C_{j_1, \cdots  j_{k-1}, i_k \cdots, i_{nc}}^{\bm{b}^1_{t_2} \cdots  \bm{b}^{k-1}_{t_2},\bm{b}^k_{t_1} \cdots, \bm{b}^{nc}_{t_1}}\left( t_2\right) 
\end{eqnarray}
\noindent For this projection, some remarks are in order:
\begin{enumerate}
\item The partial projection starts with $k=1$ and ends at $k=nc$.
\item For a given $k$, the partial projection has to be applied for all values of the indices $\left\lbrace j_1, \cdots j_k, i_{k+1} \cdots, i_{nc}\right\rbrace $.
\item  The overlap elements obtained by $n$D-multidimensional integrals, $\braket{B_{J}\left(\bm{\Lambda}_{t_2}\right) }{B_{I}\left(\bm{\Lambda}_{t_1}\right) }$, are substituted by several $1$D-integrals, \\
    $\braket{\bm{b}^k_{j_k}\left(t_2\right) }{\bm{b}^k_{i_k}\left(t_1\right) }  \equiv \braket{\bm{b}^k_{j_k}\left(\lambda_k^{(t_2)}\right) }{\bm{b}^k_{i_k}\left(\lambda_k^{(t_1)}\right) }$. 
\item These $1$D-integrals are just the overlaps, $\mathcal{S}^k\left(j_k, i_k\right) $, between the $k^{th}$ $1$D-basis sets at different time.
\end{enumerate}
The calculation of the $\mathcal{S}^k\left( j_k, i_k\right) $ overlap matrix elements are performed numerically using Gauss-Hermite quadratures:
\begin{eqnarray} \label{eq overlap_k}
\mathcal{S}^k(j_k, i_k) &=& \braket{\bm{b}^k_{j_k}\left(t_2\right) }{\bm{b}^k_{i_k}\left(t_1\right) }  \nonumber \\
&\approx& \sum_{u=1}^{nq_k} b^k_{j_k}\left( q_k^u,t_2\right)  \cdot b^k_{i_k}\left( q_k^u,t_1\right)  \cdot w_k^u
\end{eqnarray}
\noindent In Eq. \ref{eq overlap_k}, the $q_k^u$ and  the $w_k^u \left( u \in \{1 \cdots nq_k\}\right)$ represent the Gauss quadrature, $\bm{g}^{k}$, for the shifted and scaled Harmonic Oscillator basis set. In term of standard Harmonic Oscillator grid ($x_u$ and $w_u$), $q_k^u$ and $w_k^u$ are expressed as follows:
\begin{eqnarray}
q_k^u  = q^{(t)}_k + x_u/\sqrt{a^{(t)}_k}, \quad
w_k^u = w_u /\sqrt{a^{(t)}_k}
\end{eqnarray}
\noindent It is important to note that the grid used to compute $\mathcal{S}^k\left( j_k, i_k\right) $ is neither the grid associated with time $t_1$ nor the one associated with time $t_2$. Indeed, since the product of two Gaussians is also a Gaussian, this grid is an intermediate grid constructed from the time-dependent basis set parameters at times $t_1$ and $t_2$. More precisely, it can be shown that the intermediate parameters, $a^{Int}_k$ and $q^{Int}$ are:
\bea
a_{k}^{Int} &=&   \frac{a^{(t_1)}_k+a^{(t_2)}_k}{2} \\
q_{k}^{Int} &=& \frac{a^{(t_1)}_k \cdot q_{k}^{(t_1)} +a^{(t_2)}_k \cdot q_{k}^{(t_2)}}{a^{(t_1)}_k+a^{(t_2)}_k}
\ena
\noindent Remarks:  (i) we do not transform the basis sets $\bm{b}^k\left( q_k,\lambda_k^{(t_1)}\right) $ and $\bm{b}^k\left( q_k,\lambda_k^{(t_2)}\right) $ to a new basis set with the intermediate parameters, but we just compute $b^k_{j_k}\left( q_k^u,\lambda_k^{(t_2)}\right) $ and  $b^k_{i_k}\left( q_k^u,\lambda_k^{(t_1)}\right) $ on the new grid points. (ii) When, $b^{(t)}_k$ (imaginary part of $\alpha^{(t)}_k$) and $p^{(t)}_k$ are not taken into account, the computation of $\mathcal{S}^k\left( j_k, i_k\right) $ are numerically exact as soon as $nq_k=nb_k$. (iii) To reach a good accuracy, we chose $nq_k=nb_k+5$. 
\section{Appendix: 2D-retinal model} \label{2D-retinal}
\noindent The 2D-retinal model is described by a $2 \times 2$ Hamiltonian, $\hat{H}^d$, in the diabatic basis, expressed as:
\begin{equation}
\hat{H}^{d}\left(\bm{Q} \right) = \hat{T}\left(\bm{Q} \right) + 
\begin{bmatrix}
V^{d}_{11}\left(\Phi, Q \right) & V^{d}_{12}\left( Q\right)  \\
V^{d}_{21}\left( Q\right) & V^{d}_{22}\left(\Phi, Q \right)
\end{bmatrix},
\end{equation}
where the potentials for the two diabatic electronic states and their coupling are defined by:
\begin{align}
V^{d}_{11}(\Phi, Q) &= \frac{1}{2}W_1(1 - \cos \Phi) + \frac{1}{2}\omega Q^2, \\
V^{d}_{22}(\Phi, Q) &= E - \frac{1}{2}W_2(1 - \cos \Phi) + \frac{1}{2}\omega Q^2 + \kappa Q, \\
V^{d}_{12}(Q) &= V_{21}(Q) = \lambda Q.
\end{align}
Here, $m$ represents the effective mass of the reaction coordinate while $\omega$ denotes the frequency of the coupling mode. The parameter $\lambda$ corresponds to the inter-state coupling, and $W_0$, $W_1$, and $E$ are adjusted to reproduce the properties of the electronic states $S_1$ and $S_2$. The coordinate $\Phi$ represents the mode along which the molecule undergoes isomerization, and $Q$ corresponds to the active vibronic mode coupling the two electronic states. The kinetic energy operator, assumed to be the same for both diabatic electronic states, is given by:
\begin{equation}
\hat{T}\left(\bm{Q} \right)  = -\frac{1}{2m} \frac{\partial^2}{\partial \Phi^2} - \frac{\omega}{2} \frac{\partial^2}{\partial Q^2}.
\end{equation}
The values of the model parameters, determined through fitting to experimental data and theoretical calculations, are:
\begin{itemize}
\item the inverse of the effective mass, $m^{-1} = \SI{4.84e-4}{\electronvolt}$
\item the torsional barriers $W_0 = \SI{3.6}{\electronvolt}$ and $W_1 = \SI{1.09}{\electronvolt}$
\item the linear coupling $\kappa = \SI{0.1}{\electronvolt}$
\item the vibronic frequency $\omega = \SI{0.19}{\electronvolt}$
\item the inter-state coupling $\lambda = \SI{0.19}{\electronvolt}$, 
\item the offset energy  $E = \SI{2.48}{\electronvolt}$.
\end{itemize}
The basis set and initial wave packet parameters are given in the 
 table \ref{tab-retinal}.

\begin{table}[H]
	\captionsetup{width=\linewidth}
	\small
	\caption{\textit{Hagedorn basis set parameter values: $q_k^{(0)}$, $p_k^{(0)}$, and $\alpha_k^{(0)}$ (columns 2--4). Initial Gaussian wave packet parameters (Eq.~\ref{Eq-Psi0}, with $e_0 = 2$): $q_k^0$, $p_k^0$, and $\Re\left( a_k^0 \right)$ (columns 5--7)}.}
	\label{tab-retinal}
	\centering
	\begin{tabularx}{\linewidth}{@{}c *{3}{>{\centering\arraybackslash}X} *{3}{>{\centering\arraybackslash}X}@{}}
		\toprule
		& \multicolumn{3}{c}{Hagedorn and HO basis set} & \multicolumn{3}{c}{Initial wave packet} \\
		\cmidrule(lr){2-4} \cmidrule(lr){5-7}
		$k$ (coord.) 
		& $q_k^{(0)}$ & $p_k^{(0)}$ & $\alpha_k^{(0)}$ 
		& $q_k^0$ & $p_k^0$ & $\Re\left( a_k^0 \right)$ \\
		\midrule
		1 ($\Phi$) & /   & /   & /           & 0.0 & 0.0 & 60.9836267 \\
		2 ($Q$)    & 0.0 & 0.0 & 0.9202033  & 0.0 & 0.0 & 0.9202033 \\
		\bottomrule
	\end{tabularx}
	\normalsize
\end{table}
\quad \newline

\bibliographystyle{tfo}
\bibliography{Hagedorn}

\begin{thebibliography}{119}
\providecommand{\url}[1]{\texttt{#1}}
\providecommand{\urlprefix}{URL }

\bibitem{walker1994precision}
B. Walker, B. Sheehy, L. Dimauro, P. Agostini, K. Schafer and K. Kulander,
  Physical Review Letters  \textbf{73} (9), 1227--1230 (1994).

\bibitem{rudenko2004correlated}
A. Rudenko, K. Zrost, B. Feuerstein, V. DeJesus, C. Schr{\"o}ter, R. Moshammer
  and J. Ullrich,  Physical Review Letters  \textbf{93} (25), 253001 (2004).

\bibitem{engel2005calculated}
E. Engel, N. Doss, G. Harris and J. Tennyson,  Monthly Notices of the Royal
  Astronomical Society  \textbf{357} (2), 471--477 (2005).

\bibitem{pedersen2007crossed}
H. Pedersen, S. Altevogt, B. Jordon-Thaden, O. Heber, M. Rappaport, D. Schwalm,
  J. Ullrich, D. Zajfman, R. Treusch, N. Guerassimova {\em{et~al.}},  Physical
  Review Letters  \textbf{98} (22), 223202 (2007).

\bibitem{sodoga2009photodissociation}
K. Sodoga, J. Loreau, D. Lauvergnat, Y. Justum, N. Vaeck and M.
  Desouter-Lecomte,  Physical Review A  \textbf{80} (3), 033417 (2009).

\bibitem{hahn2000femtosecond}
S. Hahn and G. Stock,  Chemical Physics  \textbf{259} (2-3), 297--312 (2000).

\bibitem{hahn2000quantum}
S. Hahn and G. Stock,  the Journal of Physical Chemistry B  \textbf{104} (6),
  1146--1149 (2000).

\bibitem{marsili2020quantum}
E. Marsili, M. Olivucci, D. Lauvergnat and F. Agostini,  Journal of Chemical
  Theory and Computation  \textbf{16} (10), 6032--6048 (2020).

\bibitem{manathunga2017impact}
M. Manathunga, X. Yang, Y. Orozco-Gonzalez and M. Olivucci,  The Journal of
  Physical Chemistry Letters  \textbf{8} (20), 5222--5227 (2017).

\bibitem{Pereira2023}
A. Pereira, J. Knapik, A. Chen, D. Lauvergnat and F. Agostini,  The European
  Physical Journal Special Topics  \textbf{232} (12), 1917--1933 (2023).

\bibitem{kimura1986charge}
M. Kimura and C.D. Lin,  Physical Review A  \textbf{34} (1), 176 (1986).

\bibitem{jain1987density}
A. Jain, C.D. Lin and W. Fritsch,  Physical Review A  \textbf{36} (5), 2041
  (1987).

\bibitem{loreau2010ab}
J. Loreau, K. Sodoga, D. Lauvergnat, M. Desouter-Lecomte and N. Vaeck,
  Physical Review A  \textbf{82} (1), 012708 (2010).

\bibitem{Cigrang2025}
L.L.E. Cigrang, B.F.E. Curchod, R.A. Ingle, A. Kelly, J.R. Mannouch, D.
  Accomasso, A. Alijah, M. Barbatti, W. Chebbi, N. Do{\v{s}}li{\'{c}}, E.C.
  Eklund, S. Fernandez-Alberti, A. Freibert, L. Gonz{\'{a}}lez, G. Granucci,
  F.J. Hern{\'{a}}ndez, J. Hern{\'{a}}ndez-Rodr{\'{i}}guez, A. Jain, J.
  Jano{\v{s}}, I. Kassal, A. Kirrander, Z. Lan, H.R. Larsson, D. Lauvergnat, B.
  {Le D{\'{e}}}, Y. Lee, N.T. Maitra, S.K. Min, D. Pel{\'{a}}ez, D. Picconi, Z.
  Qiu, U. Raucci, P. Robertson, E. {Sangiogo Gil}, M. Sapunar, P.
  Sch{\"{u}}rger, P. Sinnott, S. Tretiak, A. Tikku, P. Vindel-Zandbergen, G.A.
  Worth, F. Agostini, S. G{\'{o}}mez, L.M. Ibele and A. Prlj,  The Journal of
  Physical Chemistry A   (2025).

\bibitem{Wu2025}
B. Wu, B. Li, X. He, X. Cheng, J. Ren and J. Liu,  Journal of Chemical Theory
  and Computation  \textbf{21} (8), 3775--3813 (2025).

\bibitem{miller2001semiclassical}
W.H. Miller,  The Journal of Physical Chemistry A  \textbf{105} (13),
  2942--2955 (2001).

\bibitem{liu2007real}
J. Liu and W.H. Miller,  The Journal of Chemical Physics  \textbf{126} (23),
  234110 (2007).

\bibitem{frantsuzov2003gaussian}
P. Frantsuzov, A. Neumaier and V.A. Mandelshtam,  Chemical Physics Letters
  \textbf{381} (1-2), 117--122 (2003).

\bibitem{Tully1990}
J.C. Tully,  The Journal of Chemical Physics  \textbf{93} (2), 1061--1071
  (1990).

\bibitem{Coker1995}
D.F. Coker and L. Xiao,  The Journal of Chemical Physics  \textbf{102} (1),
  496--510 (1995).

\bibitem{Kapral2016}
R. Kapral,  Chemical Physics  \textbf{481}, 77--83 (2016).

\bibitem{Agostini2016}
F. Agostini, S.K. Min, A. Abedi and E.K.U. Gross,  Journal of Chemical Theory
  and Computation  \textbf{12} (5), 2127--2143 (2016).

\bibitem{Runeson2023}
J.E. Runeson and D.E. Manolopoulos,  The Journal of Chemical Physics
  \textbf{159} (9), 104111 (2023).

\bibitem{martinez1996multi}
T.J. Martinez, M. Ben-Nun and R.D. Levine,  The Journal of Physical Chemistry
  \textbf{100} (19), 7884--7895 (1996).

\bibitem{worth2004novel}
G.A. Worth, M.A. Robb and I. Burghardt,  Faraday discussions  \textbf{127},
  307--323 (2004).

\bibitem{liu2011approach}
J. Liu and W.H. Miller,  The Journal of Chemical Physics  \textbf{134} (10),
  104102 (2011).

\bibitem{curchod2018ab}
B.F.E. Curchod and T.J. Martinez,  Chemical Reviews  \textbf{118} (7),
  3305--3336 (2018).

\bibitem{meyer1990multi}
H.D. Meyer, U. Manthe and L.S. Cederbaum,  Chemical Physics Letters
  \textbf{165} (1), 73--78 (1990).

\bibitem{Wang2002}
H. Wang and M. Thoss,  The Journal of Chemical Physics  \textbf{119} (3),
  1289--1299 (2003).

\bibitem{Manthe2008}
U. Manthe,  The Journal of Chemical Physics  \textbf{128} (16), 164116 (2008).

\bibitem{Vendrell2011}
O. Vendrell and H.D. Meyer,  The Journal of Chemical Physics  \textbf{134} (4),
  044135 (2011).

\bibitem{leforestier1991comparison}
C. Leforestier, R.H. Bisseling, C. Cerjan, M.D. Feit, R. Friesner, A.
  Gulderberg, A. Hammerich, G. Jolicard, W. Karrlein, H.D. Meyer {\em{et~al.}},
   Journal of Computational Physics  \textbf{94} (1), 59--80 (1991).

\bibitem{feit1982solution}
M.D. Feit, J.A. Fleck and A. Steiger,  Journal of Computational Physics
  \textbf{47} (3), 412--433 (1982).

\bibitem{tal1984accurate}
H. Tal-Ezer and R. Kosloff,  The Journal of Chemical Physics  \textbf{81} (9),
  3967--3971 (1984).

\bibitem{lauvergnat2007simple}
D. Lauvergnat, S. Blasco, X. Chapuisat and A. Nauts,  The Journal of chemical
  physics  \textbf{126} (20), 204103 (2007).

\bibitem{heller1976time}
E.J. Heller,  The Journal of Chemical Physics  \textbf{64} (1), 63--73 (1976).

\bibitem{huber1987generalized}
D. Huber and E.J. Heller,  The Journal of chemical physics  \textbf{87} (9),
  5302--5311 (1987).

\bibitem{drolshagen1983time}
G. Drolshagen and E.J. Heller,  The Journal of chemical physics  \textbf{79}
  (4), 2072--2082 (1983).

\bibitem{vanivcek2021ab}
J. Vaníček and T. Begušić, {Chapter 6 - Ab Initio Semiclassical Evaluation
  of Vibrationally Resolved Electronic Spectra With Thawed Gaussians in
  Molecular Spectroscopy and Quantum Dynamics} Elsevier 2021.

\bibitem{sielk2009quantum}
J. Sielk, H.F.V. Horsten, F. Krüger, R. Schneider and B. Hartke,  Physical
  Chemistry Chemical Physics  \textbf{11} (3), 463--475 (2009).

\bibitem{Choi2019}
S. Choi and J. Van{\'{i}}{\v{c}}ek,  The Journal of Chemical Physics
  \textbf{151} (23) (2019).

\bibitem{hammerich1990quantum}
A.D. Hammerich, R. Kosloff and M.A. Ratner,  Chemical Physics Letters
  \textbf{171} (1-2), 97--108 (1990).

\bibitem{meyer2003quantum}
H.D. Meyer and G.A. Worth,  Theoretical Chemistry Accounts  \textbf{109},
  251--267 (2003).

\bibitem{Pelaez2017}
D. Pel{\'{a}}ez and H.D. Meyer,  Chemical Physics  \textbf{482}, 100--105
  (2017).

\bibitem{Shi2023}
L. Shi, M. Schr{\"{o}}der, H.D. Meyer, D. Pel{\'{a}}ez, A.M. Wodtke, K.
  Golibrzuch, A.M. Sch{\"{o}}nemann, A. Kandratsenka and F. Gatti,  The Journal
  of Chemical Physics  \textbf{159} (19), 194102 (2023).

\bibitem{hagedorn1981semiclassical}
G. Hagedorn,  Annals of Physics  \textbf{135} (1), 58--70 (1981).

\bibitem{hagedorn2006ac}
G.A. Hagedorn, V. Rousse and S.W. Jilcott,  Annales Henri Poincaré
  \textbf{7}, 1065--1083 (2006).

\bibitem{hagedorn1998raising}
G. Hagedorn,  Annals of Physics  \textbf{269} (1), 77--104 (1998).

\bibitem{faou2009computing}
E. Faou, V. Gradinaru and C. Lubich,  SIAM Journal on Scientific Computing
  \textbf{31} (4), 3027--3041 (2009).

\bibitem{lasser2020computing}
C. Lasser and C. Lubich,  Acta Numerica  \textbf{29}, 229--401 (2020).

\bibitem{zhang2024single}
Z.T. Zhang and J.J. Van{\'\i}{\v{c}}ek,  The Journal of Chemical Physics
  \textbf{161} (11), 111101 (2024).

\bibitem{dietert2017invariant}
H. Dietert, J. Keller and S. Troppmann,  Journal of Mathematical Analysis and
  Applications  \textbf{450} (2), 1317--1332 (2017).

\bibitem{vanivcek2024hagedorn}
J.J.L. Van{\'{i}}{\v{c}}ek and Z.T. Zhang,  Journal of Physics A: Mathematical
  and Theoretical  \textbf{58} (8), 085303 (2025).

\bibitem{ohsawa2019hagedorn}
T. Ohsawa,  Journal of Fourier Analysis and Applications  \textbf{25},
  1513--1552 (2019).

\bibitem{hagedorn2013minimal}
G.A. Hagedorn,  Spectral Analysis, Differential Equations and Mathematical
  Physics: A Festschrift in Honor of Fritz Gesztesy's 60th Birthday
  \textbf{87}, 183--190 (2013).

\bibitem{ohsawa2013symplectic}
T. Ohsawa and M. Leok,  Journal of Physics A: Mathematical and Theoretical
  \textbf{46} (40), 405201 (2013).

\bibitem{gradinaru2014convergence}
V. Gradinaru and G.A. Hagedorn,  Numerische Mathematik  \textbf{126} (1),
  53--73 (2014).

\bibitem{lasser2014hagedorn}
C. Lasser and S. Troppmann,  Journal of Fourier Analysis and Applications
  \textbf{20}, 679--714 (2014).

\bibitem{li2014improved}
X. Li and A. Xiao,  International Journal of Modeling, Simulation, and
  Scientific Computing  \textbf{5} (04), 1450013 (2014).

\bibitem{ohsawa2015symmetry}
T. Ohsawa,  Journal of Mathematical Physics  \textbf{56} (3) (2015).

\bibitem{punovsevac2016dynamics}
P. Puno{\v{s}}evac and S.L. Robinson,  Journal of Mathematical Physics
  \textbf{57} (9), 092102 (2016).

\bibitem{ISSA_2025}
R. Issa, Wave packet propagation {F}ortran code (standard and {H}agedorn
  schemes) \url{https://github.com/asodoga/TD_Schrod_Rabiou} 2025, Accessed:
  May 02, 2025.

\bibitem{Mercouris1994}
T. Mercouris, Y. Komninos, S. Dionissopoulou and C.A. Nicolaides,  Physical
  Review A  \textbf{50}, 4109 (1994).

\bibitem{heller1981semiclassical}
E.J. Heller,  Accounts of Chemical Research  \textbf{14} (12), 368--375 (1981).

\bibitem{heller1975time}
E.J. Heller,  The Journal of Chemical Physics  \textbf{62} (4), 1544--1555
  (1975).

\bibitem{moghaddasi2023high}
R. Moghaddasi~Fereidani and J.J. Van{\'\i}{\v{c}}ek \textbf{159} (9), 094114
  (2023).

\bibitem{lasorne2011excited}
B. Lasorne, G.A. Worth and M.A. Robb,  Wiley Interdisciplinary Reviews:
  Computational Molecular Science  \textbf{1} (3), 460--475 (2011).

\bibitem{Lasorne2014}
B. Lasorne, G.A. Worth and M.A. Robb, in \emph{Molecular Quantum Dynamics: From
  Theory to Applications}, edited by Fabien Gatti  (Springer Berlin Heidelberg,
  Berlin, Heidelberg, 2014), pp. 181--211.

\bibitem{lasorne2006direct}
B. Lasorne, M.A. Robb and G.A. Worth,  Chemical Physics Letters  \textbf{432}
  (4-6), 604--609 (2006).

\bibitem{lasorne2007direct}
B. Lasorne, M.A. Robb and G.A. Worth,  Physical Chemistry Chemical Physics
  \textbf{9} (25), 3210--3227 (2007).

\bibitem{mendive2012towards}
D. Mendive-Tapia, B. Lasorne, G.A. Worth, M.A. Robb and M.J. Bearpark,  The
  Journal of Chemical Physics  \textbf{137} (22), 22A513 (2012).

\bibitem{allan2010straightforward}
C.S. Allan, B. Lasorne, G.A. Worth and M.A. Robb,  The Journal of Physical
  Chemistry A  \textbf{114} (33), 8713--8729 (2010).

\bibitem{araujo2009molecular}
M. Araújo, B. Lasorne, A.L. Magalhães, G.A. Worth, M.J. Bearpark and M.A.
  Robb,  The Journal of Chemical Physics  \textbf{131} (14), 144301 (2009).

\bibitem{lasorne2008controlling}
B. Lasorne, M.J. Bearpark, M.A. Robb and G.A. Worth,  The Journal of Physical
  Chemistry A  \textbf{112} (50), 13017--13027 (2008).

\bibitem{araujo2010controlling}
M. Araujo, B. Lasorne, A.L. Magalhaes, M.J. Bearpark and M.A. Robb,  The
  Journal of Physical Chemistry A  \textbf{114} (45), 12016--12020 (2010).

\bibitem{richings2015quantum}
G.W. Richings, I. Polyak, K.E. Spinlove, G.A. Worth, I. Burghardt and B.
  Lasorne,  International Reviews in Physical Chemistry  \textbf{34} (2),
  269--308 (2015).

\bibitem{beck2000multiconfiguration}
M.H. Beck, A. J{\"a}ckle, G.A. Worth and H.D. Meyer,  Physics Reports
  \textbf{324} (1), 1--105 (2000).

\bibitem{MCTDHWebsite}
H.D. Meyer, U. Manthe and L. Cederbaum, The Multiconfiguration Time-Dependent
  Hartree (MCTDH) Method. https://www.pci.uni-heidelberg.de/tc/mctdh.html.

\bibitem{Nest2002}
M. Nest and H.D. Meyer,  The Journal of Chemical Physics  \textbf{117} (23),
  10499 (2002).

\bibitem{mclachlan1964time}
A. McLachlan and M. Ball,  Reviews of Modern Physics  \textbf{36} (3), 844
  (1964).

\bibitem{Wilson1955a}
E.B. Wilson, J.C. Decius and P.C. Cross, \emph{{Molecular Vibrations: The
  Theory of Infrared and Raman Vibrational Spectra}}, mcgraw-hil ed. New York.

\bibitem{Meyer1968}
R. Meyer and H.H. G{\"{u}}nthard,  The Journal of Chemical Physics  \textbf{49}
  (4), 1510--1520 (1968).

\bibitem{Chapuisat1991}
X. Chapuisat, A. Nauts and J.P. Brunet,  Molecular Physics  \textbf{72} (1),
  1--31 (1991).

\bibitem{Chapuisat1992a}
X. Chapuisat and C. Iung,  Physical Review A  \textbf{45} (9), 6217--6235
  (1992).

\bibitem{Wang2000a}
X.G. Wang and T. Carrington,  The Journal of Chemical Physics  \textbf{113}
  (17), 7097--7101 (2000).

\bibitem{Gatti2001a}
F. Gatti, C. Mu{\~{n}}oz and C. Iung,  Journal of Chemical Physics
  \textbf{114} (19), 8275--8282 (2001).

\bibitem{Ndong2012}
M. Ndong, L. Joubert-Doriol, H.D. Meyer, A. Nauts, F. Gatti and D. Lauvergnat,
  The Journal of Chemical Physics  \textbf{136} (3), 034107 (2012).

\bibitem{Ndong2013}
M. Ndong, A. Nauts, L. Joubert-Doriol, H.D. Meyer, F. Gatti and D. Lauvergnat,
  The Journal of Chemical Physics  \textbf{139} (20), 204107 (2013).

\bibitem{Meyer1979}
R. Meyer,  Journal of Molecular Spectroscopy  \textbf{76}, 266--300 (1979).

\bibitem{Harthcock1982}
M. Harthcock and J. Laane,  Journal of Molecular Spectroscopy  \textbf{324},
  300--324 (1982).

\bibitem{Senent1995}
M.L. Senent, D.C. Moule and Y.G. Smeyers,  Journal of Molecular Spectroscopy
  \textbf{372}, 257--266 (1995).

\bibitem{Luckhaus2000}
D. Luckhaus,  The Journal of Chemical Physics  \textbf{113} (4), 1329--1347
  (2000).

\bibitem{Lauvergnat2002}
D. Lauvergnat and A. Nauts,  The Journal of Chemical Physics  \textbf{116}
  (19), 8560--8570 (2002).

\bibitem{Matyus2009}
E. Mátyus, G. Czakó and A.G. Császár,  The Journal of chemical physics
  \textbf{130} (13), 134112 (2009).

\bibitem{Marsili2022}
E. Marsili, F. Agostini, A. Nauts and D. Lauvergnat,  Philosophical
  Transactions of the Royal Society A: Mathematical, Physical and Engineering
  Sciences  \textbf{380} (2223), 20200388 (2022).

\bibitem{gottlieb1977numerical}
D. Gottlieb and S.A. Orszag, Numerical analysis of spectral methods: theory and
  applications SIAM 1977.

\bibitem{light2000discrete}
J.C. Light and T. Carrington,  Advances in Chemical Physics  \textbf{114},
  263--310 (2000).

\bibitem{friesner1986solution}
R.A. Friesner,  The Journal of Chemical Physics  \textbf{85} (3), 1462--1468
  (1986).

\bibitem{szalay2012variational}
V. Szalay and P. {\'A}d{\'a}m,  The Journal of Chemical Physics  \textbf{137}
  (6), 064118 (2012).

\bibitem{yu2005coherent}
H.G. Yu,  The Journal of Chemical Physics  \textbf{122} (16), 164107 (2005).

\bibitem{gatti1999fully}
F. Gatti, C. Iung, C. Leforestier and X. Chapuisat,  The Journal of chemical
  physics  \textbf{111} (16), 7236--7243 (1999).

\bibitem{Light2000}
J.C. Light and T. {Carrington Jr},  Advances in Chemical Physics  \textbf{114},
  263--310 (2000).

\bibitem{johnson2017primary}
P.J.M. Johnson, M.H. Farag, A. Halpin, T. Morizumi, V.I. Prokhorenko, J.
  Knoester, T.L.C. Jansen, O.P. Ernst and R.J.D. Miller,  The Journal of
  Physical Chemistry B  \textbf{121} (16), 4040--4047 (2017).

\bibitem{seidner1994microscopic}
L. Seidner and W. Domcke,  Chemical Physics  \textbf{186} (1), 27--40 (1994).

\bibitem{domcke1997theory}
W. Domcke and G. Stock,  Advances in Chemical Physics  \textbf{100}, 1--169
  (1997).

\bibitem{balzer2003mechanism}
B. Balzer, S. Hahn and G. Stock,  Chemical Physics Letters  \textbf{379} (3-4),
  351--358 (2003).

\bibitem{Ben-Nun1998}
M. Ben-Nun and T.J. Martínez,  The Journal of Chemical Physics  \textbf{108}
  (17), 7244--7257 (1998).

\bibitem{Gradinaru2024}
V. Gradinaru and O. Rietmann,  Journal of Computational Physics  \textbf{509}
  (April), 113029 (2024).

\bibitem{Dalibard1992}
J. Dalibard, Y. Castin and K. M{\o}lmer,  Physical Review Letters  \textbf{68}
  (5), 580--583 (1992).

\bibitem{Gerdts1997}
T. Gerdts and U. Manthe,  Journal of Chemical Physics  \textbf{106} (8),
  3017--3023 (1997).

\bibitem{Smolyak1963a}
S.A. Smolyak,  Soviet Mathematics Doklady  \textbf{4}, 240 (1963).

\bibitem{Gradinaru2008}
V. Gradinaru,  SIAM Journal on Numerical Analysis  \textbf{46} (1), 103--123
  (2008).

\bibitem{Xie-Chen2024}
Y. Xie, Y. Yang, X. Zhu, A. Chen and B. Gu,  Journal of Chemical Theory and
  Computation  \textbf{20} (21), 9512--9521 (2024).

\bibitem{Avila2009}
G. Avila and T. Carrington,  The Journal of chemical physics  \textbf{131}
  (17), 174103 (2009).

\bibitem{Avila2017a}
G. Avila and T. Carrington,  Chemical Physics  \textbf{482}, 3--8 (2017).

\bibitem{Lauvergnat2013}
D. Lauvergnat and A. Nauts,  Spectrochimica Acta Part A: Molecular and
  Biomolecular Spectroscopy  \textbf{119}, 18--25 (2014).

\bibitem{Avila2019}
G. Avila and E. M{\'{a}}tyus,  The Journal of Chemical Physics  \textbf{150}
  (17), 174107 (2019).

\bibitem{Chen2022}
A. Chen, D.M. Benoit, Y. Scribano, A. Nauts and D. Lauvergnat,  Journal of
  Chemical Theory and Computation  \textbf{18} (7), 4366--4372 (2022).

\bibitem{Lauvergnat2023}
D. Lauvergnat and A. Nauts,  ChemPhysChem  \textbf{24} (21), e202300501 (2023).

\end{thebibliography}

\end{document}